\documentclass[a4papper,10pt]{article}
\usepackage{amssymb}
\usepackage{graphicx}
\usepackage{ifpdf}
\begin{document}
\def\ctr#1{\hfil $\,\,\,#1\,\,\,$ \hfil}
\def\tstrut{\vrule height 2.7ex depth 1.0ex width 0pt}
\def\mystrut{\vrule height 3.7ex depth 1.6ex width 0pt}
\def \inparg{\leftskip = 40pt\rightskip = 40pt}
\def \outparg{\leftskip = 0 pt\rightskip = 0pt}
\def\lf{16\pi^2}
\def\beqn{\begin{eqnarray}}
\def\eeqn{\end{eqnarray}}
\def\llf{(16\pi^2)^2}
\def\lllf{(16\pi^2)^3}
\def\llllf{(16\pi^2)^4}
\def\tr{\, {\rm tr}}
\def\Ttil{\widetilde{T}}
\def\Btil{\widetilde{B}}
\def\Etil{\widetilde{E}}
\def\ga{\gamma}
\def\b{\textbf}
\def\frak{\frac}
\def\ttil{\widetilde{t}}
\def\util{\widetilde{u}}
 \def\btil{\widetilde{b}}
\def\mutil{\widetilde{\mu}}
\def\dtil{\widetilde{d}}
\def\ctil{\widetilde{c}}
\def\stil{\widetilde{s}}
\def\tautil{\widetilde{\tau}}
\def\etil{\widetilde{e}}
\def\nutil{\widetilde{\nu}}
\def\gtil{\widetilde{g}}
\def\bigpound{ hbox{\bigit\$}} 
\def \q{\quad}
\def\lf{16\pi^2}
\def\llf{(16\pi^2)^2}
\def\lllf{(16\pi^2)^3}
\def \qq{\qquad}
\def \b{\bigskip}
\def\GeV{{\rm GeV}}
 \def\ha{{1\over2}}

\def\half{{\textstyle{1\over2}}} 
\def\frak#1#2{{\textstyle{{#1}\over{#2}}}}
\def\frakk#1#2{{{#1}\over{#2}}}
\def\pa{\partial}
\def\Scal{{\cal S}}

\def\cosec{\hbox{cosec }}
\def\semi{;\hfil\break}
\def\NSVZ{{\rm NSVZ}}
\def\DRED{{\rm DRED}}
\def\DREDp{{\rm
DRED}'}
\def\ga{\gamma}
\def\de{\delta}
\def\ep{\epsilon}
\def \la{\lambda}
\def \La{\Lambda}
\def \th{\theta}
\def\al{\alpha}

\def \bfa{{\bf a}}
\def \bfb{{\bf b}}
\def \bfc{{\bf c}}
\def \bfi{{\bf i}}
\def \bfj{{\bf j}}
\def \bfk{{\bf k}}
\def \bfn{{\bf n}}
\def \bfp{{\bf p}}
\def \bfr{{\bf r}}

\def \bfA{{\bf A}}
\def \bfB{{\bf B}}
\def \bfE{{\bf E}}
\def \bfF{{\bf F}}
\def \bfJ{{\bf J}}
\def \bfL{{\bf L}}
\def \bfR{{\bf R}}
\def \bfT{{\bf T}}
\def \bfV{{\bf V}}

\def\bfAdot{\dot\bfA}

\def\alphadot{\dot\alpha}
\def\betadot{\dot\beta}
\def\thdot{\dot\theta}
\def\thddot{\ddot\theta}
\def\omegadot{\dot\omega}
\def\omegaddot{\ddot\omega}
\def\onedot{\dot 1}
\def\twodot{\dot 2}
\def\pdot{\dot p}
\def\pddot{\ddot p}
\def\qdot{\dot q}
\def\qddot{\ddot q}
\def\rdot{\dot r}
\def\rddot{\ddot r}
\def\bfrdot{{\bf{\dot r}}}
\def\bfrddot{{\bf{\ddot r}}}
\def\xdot{\dot x}
\def\xddot{\ddot x}
\def\ydot{\dot y}
\def\yddot{\ddot y}
\def\zdot{\dot z}
\def\zddot{\ddot z}

\def\psib{\overline{\psi}}
\def\betab{\overline{\beta}}
\def\epb{\overline{\epsilon}}
\def\lab{\overline{\lambda}}
\def\thb{\overline{\theta}}
\def\chib{\overline{\chi}}
\def\taub{{\overline{\tau}}}
\def\phib{\overline{\phi}}
\def\Phib{\overline\Phi}
\def\Pib{\overline\Pi}
\def\sigmat{\sigma^{\mu}_{\alpha\alphadot}}
\def\sigmab{\overline{\sigma}}
\def\sigmad{\sigmab^{\mu\alphadot\alpha}}
\def\xib{\overline{\xi}}
\def\nub{\overline{\nu}}
\def\TeV{{\rm TeV}}
\def\GeV{{\rm GeV}}

\def\tautilde{\tilde\tau}
\def\chitilde{\tilde\chi}
\def\nutilde{\tilde\nu}
\def\gatilde{\tilde\ga}

\def\btilde{\tilde b}
\def\dtilde{\tilde d}
\def\etilde{\tilde e}
\def\gtilde{\tilde g}
\def\mtilde{\tilde m}
\def\ttilde{\tilde t}
\def\utilde{\tilde u}
\def\ytilde{\tilde y}

\def\Atilde{\tilde A}
\def\Btilde{\tilde B}
\def\Ctilde{\tilde C}
\def\Dtilde{\tilde D}
\def\Etilde{\tilde E}
\def\Ttilde{\tilde T}
\def\Xtilde{\tilde X}
\def\Ytilde{\tilde Y}
\def\bbar{{\overline{b}}}
\def\dbar{{\overline{d}}}
\def\ebar{{\overline{e}}}
\def\fbar{{\overline{f}}}
\def\gbar{{\overline{g}}}
\def\jbar{{\overline{j}}}
\def\mbar{{\overline{m}}}
\def\qbar{{\overline{q}}}
\def\tbar{{\overline{t}}}
\def\ubar{{\overline{u}}}
\def\ybar{{\overline{y}}}
\def\Bbar{{\overline{B}}}
\def\Dbar{{\overline{D}}}
\def\Ebar{{\overline{E}}}
\def\Hbar{{\overline{H}}}
\def\Jbar{{\overline{J}}}
\def\Qbar{{\overline{Q}}}
\def\Qb{\overline{Q}}
\def\Ubar{{\overline{U}}}
\def\Wbar{{\overline{W}}}
\def\Zbar{{\overline{Z}}}
\def\fivebar{{\overline{5}}}
\def\tenbar{{\overline{5}}}
\def\threebar{{\overline{3}}}
\def\phibar{{\overline{\phi}}}
\def\taubar{{\overline{\tau}}}

\def\msbar{{\overline{\rm MS}}}
\def\drbar{{\overline{\rm DR}}}
\def\gahat{\hat{\gamma}}
\def\lahat{\hat{\lambda}}
\def\ephat{\hat{\epsilon}}
\def\ghat{\hat{g}}
\def\Bhat{\hat{B}}
\def\Khat{\hat{K}}

\def\lt{\lambda_{{}_T}}
\def\mt{m_{{}_T}}
\def\leff{{\cal L}_{eff}}
\def\tc{T_c}
\def\vev#1{\mathopen\langle #1\mathclose\rangle }
\def\Dslash{D\!\!\!\! /}
\def\dslash{\pa \!\!\! /}
\def\kslash{k\!\!\! /}
\def\pslash{p\!\!\! /}

\def\sy{supersymmetry}
\def\sic{supersymmetric}
\def\sa{supergravity}
\def\ssm{supersymmetric standard model}
\def\sm{standard model}
\def\ssb{spontaneous symmetry breaking}
\def\smgroup{$SU_3\otimes\ SU_2\otimes\ U_1$}
\def\stw{\sin^2\th_W}

\def\app{{Acta Phys.\ Pol.\ }{\bf B}}
\def\anp{Ann.\ Phys.\ }
\def\cmp{Comm.\ Math.\ Phys.\ }
\def\fortphys{{Fort.\ Phys.\ }{\bf A}}
\def\ijmpa{{Int.\ J.\ Mod.\ Phys.\ }{\bf A}}
\def\jetp{JETP\ }
\def\jetpl{JETP Lett.\ }
\def\jmp{J.\ Math.\ Phys.\ }
\def\mpla{{Mod.\ Phys.\ Lett.\ }{\bf A}}
\def\nc{Nuovo Cimento\ }
\def\npb{{Nucl.\ Phys.\ }{\bf B}}
\def\physrep{Phys.\ Reports\ }
\def\plb{{Phys.\ Lett.\ }{\bf B}}
\def\pnas{Proc.\ Natl.\ Acad.\ Sci.\ (U.S.)\ }
\def\pr{Phys.\ Rev.\ }
\def\prd{{Phys.\ Rev.\ }{\bf D}}
\def\prl{Phys.\ Rev.\ Lett.\ }
\def\ptp{Prog.\ Th.\ Phys.\ }
\def\sjnp{Sov.\ J.\ Nucl.\ Phys.\ }
\def\tmp{Theor.\ Math.\ Phys.\ }
\def\pw{Part.\ World\ }
\def\zpc{Z.\ Phys.\ {\bf C}}

\def\dfx{{{df}\over{dx}}}

\def\dyx{{{dy}\over{dx}}}
\def\ddyx{{{d^2 y}\over{dx^2}}}
\def\dddyx{{{d^3 y}\over{dx^3}}}
\def\pux{{{\pa u}\over{\pa x}}}
\def\puy{{{\pa u}\over{\pa y}}}
\def\pvx{{{\pa v}\over{\pa x}}}
\def\pvy{{{\pa v}\over{\pa y}}}
\def\pzx{{{\pa z}\over{\pa x}}}
\def\pzy{{{\pa z}\over{\pa y}}}

\def\msbar{{\overline{\rm MS}}}
\def\drbar{{\overline{\rm DR}}}

\catcode`\@=11
\def\and{\char`\&}
\def\TeV{{\rm TeV}}
\def\GeV{{\rm GeV}}
\def\eV{{\rm eV}}

\vskip .3in
\centerline{\large{\textbf{ Two-loop R-parity violating  Renormalisation Group Equations for non-standard soft   }}}
\medskip
\centerline{\large{\textbf{ supersymmetry breaking in the context of the MSSM }}}
\vskip .3in
\centerline{\bf  A.F.~Kord and S.~Hosseini}
\bigskip
\centerline{\small{\it {Department of Physics,Sabzevar Tarbiat Moallem University ,
      P.O.Box 397, Sabzevar, Iran}}}
\small{\it{E-mail:farzaneh@sttu.ac.ir}}
\vskip .3in

\begin{abstract}
We present the  two-loop $\beta$-functions for non-standard soft supersymmetry-breaking couplings including non-standard R-parity violating soft terms in the Minimal Supersymmetric Standard Model

\end{abstract}

\section{Introduction}
 It is known the MSSM is a gauge supersymmetric
extension of the standard model, with the addition of a number of dimension 2 and dimension
3 supersymmetry-breaking mass and interaction terms.The model contains
many new couplings, the soft SUSY breaking interactions, which are arbitrary in the
low energy effective theory, and the reason for softly breaking terms are to avoid   unwanted quadratic divergences. In order to reduce the huge soft susy parameters very  specific scenarios have been   made about physics at energies much higher than accessible energy such as mSUGRA and CMSSM. They require an $N=1$ supergravity theory at the Planck scale with a superpotential of a very special and poorly motivated form. To obtain experimental predictions, one should extrapolate the values of these
couplings to the weak scale by using the renormalisation group equations.

However a class of low-energy supersymmetric models can be constructed by imposing some quite mild assumptions about the structure of the low-energy theory. The high-energy origin of the models is irrelevant to this discussion. It is required that the following hold at the weak scale. First, the theory has minimal particle content consistent with explaining the observed particles and being supersymmetric. This is the strongest of  assumptions and is a reasonable starting point. Second, the theory must have no quadratic divergences. The absence of quadratic divergences is a major motivation for low-energy supersymmetry, and one should only  allow all supersymmetry-breaking operators  they do not cause quadratic divergences . To avoid quadratic divergences, one needs only to prohibit all dimension-four supersymmetry breaking interactions. To see this, notice that supersymmetry breaking is now accompanied by a mass parameter $m$, so that a quadratic divergence in an operator would have a coefficient proportional to $\Lambda ^2 m$, where $\Lambda$ is a cut-off scale. Only dimension-one operators could have such a coefficient. In theories with no scalars which are singlets under all symmetries of the theory, such an operator cannot occur.

An advantage of the model which is based on the above assumptions is that the parameter $B$ is no longer dependent on the $\mu $-parameter. We know in the CMSSM the $B$ parameter is set proportional to $\mu$, and a nonzero $B$ is required for suitable electroweak symmetry breaking. Therefore, $\mu$ is required to be roughly of order $10^{2}$ or $10^{3} \GeV $ , in order to allow a Higgs vev of order 174$\GeV $. It is hard to understand why this parameter should be so small, and the same size as terms in the supersymmetry-breaking potential.
However it is possible to make a model (based on the above assumption), and in  this model  the parameter $\mu $ is no longer required in the supersymmetric potential because the general breaking potential permits a $B$ term which is in principle unrelated to the parameter $\mu$, so it is possible to set $\mu$ to zero Ref~\cite{a2}.

It is well known that the
MSSM is not, in fact, the most general renormalisable field theory consistent with the
requirements of gauge invariance and naturalness; the unbroken theory is augmented by
a discrete symmetry (R-parity) to forbid a set of baryon-number and lepton-number violating
interactions, and the supersymmetry-breaking sector omits both R-parity violating
soft terms and a set of "non-standard" (NS) soft breaking terms. There is a large literature
on the effect of R-parity violation; a recent analysis (with "standard" soft-breaking
terms) and references appears in Refs.~\cite{a3,a5,a6,a7}; for earlier relevant work see in particular Ref~\cite{a8}. The
need to consider NS terms in a model independent analysis was stressed in Ref.~\cite{a9}; for a
discussion of the NS terms both in general and in the MSSM context see  Refs.~\cite{a10,a11,a12};and
for model-building applications see for example Refs.~\cite{a13,a14}. For application of NS R-parity
violating terms to leptogenesis, see Ref.~\cite{a15}.

In a previous paper Ref.~\cite{a12} we gave the one-loop $\beta$-functions for non-standard parameters  in the MSSM context. In this paper we extend the  results to two loop corrections in the MSSM context.

\section{The new soft breaking terms}
   The minimal supersymmetric standard model (MSSM)  consists of a supersymmetric extension of the standard model, with the addition of a number of dimension 2 and dimension 3 supersymmetry-breaking mass and interaction terms. It became accepted when it was established that such a configuration is a natural consequence of supergravity when supersymmetry is broken in a hidden sector. However in Ref~\cite{a10}
  it has been shown there are  the new non-standard soft supersymmetry breaking terms which  have no quadratic divergences. These terms are allowed according to the more general philosophy explained in the previous section.
    Now we review their results for a general $N=1$ theory. We know the typical Lagrangian for  the MSSM consists of two parts

\beqn
\mathcal{L}=\mathcal{L}_{SUSY}+\mathcal{L}_{SOFT},
\eeqn

where  $\mathcal{L}_{SUSY}$ is the Lagrangian for the supersymmetric gauge theory, containing the gauge multiplet($A_{\mu}, \lambda$ that is the gaugino) and  a matter multiplet ( the spin-zero field $\phi _i$ and the spin-$\frak{1}{2}$ fields $\psi _i $). We assume a superpotential of the form
\beqn
W=\frak{1}{6}Y^{ijk}\Phi _i \Phi _j \Phi _k .
\eeqn
A renormalisable superpotential  also contains quadratic and linear terms, but we assume there are no gauge singlet fields so there is no linear term, also we assume that   an explicit quadratic  term  is not needed   because such a term will be included as a special case from the new soft breaking terms Ref~\cite{a10,a12}. A general soft breaking Lagrangian that prevents  quadratic divergences is given by:
\beqn
££\mathcal{L}_{SOFT}= (m^2)^j_i\phi ^i\phi _j+(\frak{1}{6}h^{ijk}\phi _i\phi _j\phi _k +\frak{1}{2}b^{ij}\phi _i\phi _j+\frak{1}{2}M\lambda \lambda +h.c.) +\mathcal{L}^{new}_{Soft},
\eeqn
where $\mathcal{L}^{new}_{Soft}$ introduces further possible dimension 3 terms in the case of a wide range of theories which preserve naturalness. It is given by:
\beqn
\mathcal{L}^{new}_{Soft}=\frak{1}{2}r^{jk}_{i}\phi ^i \phi _j \phi _k +\frak{1}{2} m_F^{ij}\psi _i \psi _j +m_A^{ia}\psi _i \lambda _a +h.c.
\eeqn
The $m_A$ term is only possible in the presence of adjoint matter fields~\cite{a16}.
 It is not an aspect of the MSSM, but often encountered in GUTs.   Matter fermion mass terms ($m _F$ terms) and  $r _i ^{jk}$ terms can still  guarantee
  absence of quadratic divergences if they satisfy~\cite{a17}.
\beqn
r_i^{ij}-(m_F)_{ik}Y^{ijk}=0
\eeqn
A solution to Eq (5.5) comes from modifying the superpotential $W$ as follows:
\beqn
W\to W+\frak{1}{2}(m_F)_{ij}\psi ^i \psi ^j
\eeqn
since this new superpotential gives the appropriate fermion mass and interaction. There are other possibilities, such as,
 \beqn
r_i^{ij}=(m_F)_{ik}=0 \ \ \ (\mbox{but} \   \  r_i^{jk}\neq 0).
\eeqn
Of course if there is no gauge singlet chiral superfields then $r_i^{ij}=(m_F)_{ik}Y^{ijk}=0$ holds automatically. Therefore, in most cases in the MSSM we can keep these new soft breaking terms.

\subsection{Non-standard terms in the MSSM}
Now we present the case of the MSSM in the most general possible softly-broken version of the MSSM incorporating both R-parity violating (RPV) and non-standard (NS) terms.  We include all possible soft supersymmetry breaking terms consistent with gauge invariance, which split the masses and couplings of particles and superpartners, but which do not remove the supersymmetric protection against large radiative corrections to scalar masses. It is interesting that, as we shall see, with the generalisation to the RPV case the connection between the NS terms and cubic scalar interactions involving supersymmetric mass terms is not universal.

We review the MSSM  Lagrangian   including all terms which avoid  quadratic divergences. The superpotential is defined by
\beqn
&&W=W_R+W_{NR},
\eeqn
where $W_R$ and $W_{NR}$ have been  given  by Eq (4.2) and (4.3). However, we classify  these parts again, and omit possible mass terms $H_1H_2$ and $ LH_2$ because they are the consequent terms of non- standard soft terms. In this new structure  the $W_R$ and $W_{NR}$ terms are given by
 \beqn
W_R&=&  Y_u Q  U^c H_2 +   Y_d Q D^c  H_1 +  Y_{e} L  E^c  H_1 \\
 W_{NR}& =& \frak{1}{2} (\Lambda_E) E^c L L +  \frak{1}{2}(\Lambda_U) U^c U^c D^c
+ (\Lambda_D) D^c L Q.
\eeqn
where generation $(i,j\cdots)$, $SU_2 (a,b\cdots)$,
and $SU_3 (\alpha, \beta \cdots)$  indices are
contracted in ``natural'' fashion from left to right, thus for example
\beqn
 \Lambda_D d^c L Q
\equiv  \epsilon_{ab}(\Lambda_D)^{ijk} (d^c)_{i\alpha} L^a_{j} Q^{b\alpha}_k
.
\eeqn
We show complex conjugation by lowering the indices, thus $(Y_u)_{ij} = (Y^*_u)^{ij}$.

The complete soft-breaking terms are given by
\beqn
&&L_1 = \sum_{\phi}
m_{\phi}^2\phi^*\phi + \left[m_3^2 H_1
H_2 + \sum_{i=1}^3\half M_i\lambda_i\lambda_i  + {\rm h.c. }\right]\cr
&&+ \left[ h_u\widetilde{Q} \widetilde{ U}^c H_2 +  h_d \widetilde{Q}\widetilde{D}^c  H_1 +  h_{e} \widetilde{L}\widetilde {E}^c  H_1
+ {\rm h.c. }\right],\\
&&L_2 = m_R^2   H_1^* \widetilde{L} + m_K^2 \widetilde{L} H_2 + \frak{1}{2} h_E \widetilde{E}^c\widetilde{ L}\widetilde{ L}
+  \frak{1}{2} h_U \widetilde{U}^c \widetilde{D}^c \widetilde{D}^c \cr &&
+ h_D \widetilde{D}^c \widetilde{L}\widetilde{ Q}  + {\rm h.c.,}\\
&&L_3=  m_4 \psi_{H_1}\psi_{H_2} +
R_5 H_2^*\widetilde{ L}\widetilde{ E}^c +  R_7  H_2^* \widetilde{ Q}\widetilde{D}^c  + \cr&& R_9 H_1^*\widetilde{ Q}\widetilde{ U}^c + {\rm h.c.,}\\
&&L_4 = m_r  \psi_{L}\psi_{H_2} +R_1 \widetilde{L}^* \widetilde{Q} \widetilde{U}^c +  R_2 H_1 H_2^* \widetilde{E}^c
+ R_3 \widetilde{U}^c\widetilde{E}^c \widetilde{D}^{c*} \cr
&&+ \frak{1}{2}R_4\widetilde{ Q}\widetilde{ Q}\widetilde {D}^{c*} + {\rm h.c.},
\eeqn

where $L_1$ corresponds to SRPC(Standard R-parity Conserving), $L_2$ indicates SRPV(Standard R-parity Violating), $L_3$ shows NSRPC(Non-Standard R-parity Conserving) and $L_4$ corresponds to NSRPV(Non-Standard R-parity Violating) terms. These terms come from ``pure '' trilinears $(\phi\phi\phi)$, ``mixed'' trilinears $(\phi \phi \phi ^*)$, scalar $(\phi ^* \phi)$ masses, a $ \phi \phi $ Higgs-boson mass-mixing, and $\psi \psi $ terms in Eq (5.4).

We have separated the soft terms into Eqs. (5.38) and (5.39) because  each group of couplings are not involved in the $\beta$-functions of the other group of  couplings. In the Lagrangian we have two interaction terms $(R_{3, 4})$ which cannot be generated by supersymmetric mass terms. These terms violate baryon and lepton numbers$(L, B)$, so the effects of these two terms may be comparable to $\Lambda _D, \Lambda _E$ and $ \Lambda _U $.

The full   one-loop $\beta $-functions including NS soft terms and RPV terms have been presented in ~\cite{a12}. The full three loop $\beta$-functions including soft standard $\beta$-functions have been
presented in~\cite{a18}. The two loop gauge $\beta$  functions and anomalous dimensions in the
R-parity violating (RPV) case have been calculated in~\cite{a19}. In particular, Ref ~\cite{a7} contained a
complete set of one-loop $\beta$-functions for RPV parameters. The full two loop $\beta$-functions for the RPV couplings have been presented in ~\cite{a3,a20}. In this paper, we shall write down explicitly all possible R-parity violating
terms in the framework of the Minimal Supersymmetric Standard Model, assuming the most general breaking of
 the MSSM incorporating both R-parity violating (RPV) and non-standard (NS) terms, and then   compute  results of two loop $\beta$-functions for all non-standard (NS) parameters of the MSSM  in the most general possible softly-broken version of the MSSM. Here we give the full   two-loop $\beta $-functions including NS soft terms and RPV terms.

The two loop-$\beta$-functions for $R_{1,\cdot 9}$ are given as follows:

\begin{eqnarray}
(16\pi^2)^2(\beta^{(2)}_{R_1})^{jk}_i&=&
-2(Y_d)^{il}(Y_e)_{km}C_{e^c}(R_3)^{jm}_l-2(\Lambda_D)^{lpi}(\Lambda_E)_{mpk}C_{e^c}(R_3)^{jm}_l
\nonumber\\&&+2(\Lambda_D)^{pli}(\Lambda_D)_{pkm}C_Q(R_1)^{mj}_l-2(Y_d)^{ip}(\Lambda_D)_{pkm}C_Q(R_9)^{mj}
  \nonumber\\&&+2(Y_u)^{jp}(\Lambda_D)_{mkp}C_{d_c}(R_7)^{im}-4(\Lambda_u)^{jlp}(\Lambda_D)_{pkm}C_Q(R_4)^{im}_l
             \nonumber\\&&+(\Lambda_D)^{npi}(\Lambda_E)_{mpk}(R_3)^{jm}_nC_{d_c}-2(Y_d)^{in}(Y_e)_{km}(R_3)^{jm}_nC_{d_c}
            \nonumber\\&&+2(\Lambda_D)^{pni}(\Lambda_D)_{pkm}(R_1)^{mj}_nC_L-2(Y_d)^{ip}(\Lambda_D)_{pkm}(R_9)^{mj}C_{H_1}
    \nonumber\\&&-2(Y_u)^{pj}(\Lambda_D)_{mkp}(R_7)^{im}C_{H_2}+2(\Lambda_u)^{jnp}(\Lambda_D)_{pkm}(R_4)^{im}_nC_{d_c}
\nonumber\\&&+4(Y_u)^{im}(Y_e)_{kn}(\Lambda_D)_{mpq}(Y_e)^{pn}(R_9)^{qj}+4(Y_d)^{im}(Y_e)_{kn}(\Lambda_D)_{mpq}(\Lambda_E)^{nrp}(R_1)^{qj}_r \nonumber\\&&+4(Y_d)^{im}(Y_e)_{kn}(Y_d)_{qm}(Y_e)^{rn}(R_1)^{qj}_r-2(\Lambda_D)^{lmi}(\Lambda_D)_{lkn}(\Lambda_D)_{pmq}(\Lambda_D)^{prn}(R_1)^{qj}_r
\nonumber\\&&-2(Y_d)^{il}(\Lambda_D)_{lkn}(Y_d)_{qp}(\Lambda_D)^{prn}(R_1)^{qj}_r+4(\Lambda_D)^{mli}(\Lambda_E)_{nlk}(\Lambda_D)_{mpq}(\Lambda_E)^{nrp}(R_1)^{qj}_r
\nonumber\\&&-4(\Lambda_D)^{mli}(\Lambda_E)_{nlk}(Y_d)_{qm}(Y_e)^{rn}(R_1)^{qj}_r+2(\Lambda_D)^{lmi}(\Lambda_D)_{lkn}(\Lambda_D)_{pmq}(Y_d)^{np}(R_9)^{qj}
\nonumber\\&&+2(\Lambda_D)^{lmi}(\Lambda_D)_{lkn}(Y_e)_{mq}(Y_d)^{nr}(R_3)^{jq}_r+2(\Lambda_D)^{lmi}(\Lambda_D)_{lkn}(\Lambda_E)_{qpm}(\Lambda_D)^{rpn}(R_3)^{jq}_r
\nonumber\\&&-2(Y_d)^{il}(\Lambda_D)_{lkn}(Y_e)_{pq}(\Lambda_D)^{rpn}(R_3)^{jq}_r+2(Y_d)^{il}(\Lambda_D)_{lkn}(Y_d)_{qp}(Y_d)^{np}(R_9)^{qj}
\nonumber\\&&-4(Y_u)^{lj}(\Lambda_D)_{nkl}(Y_u)_{qp}(\Lambda_U)^{prn}(R_4)^{iq}_r-4(\Lambda_U)^{jml}(\Lambda_D)_{lkn}(Y_d)_{qm}(Y_d)^{nr}(R_4)^{iq}_r
\nonumber\\&&-4(\Lambda_U)^{jml}(\Lambda_D)_{lkn}(\Lambda_D)_{mpq}(\Lambda_D)^{rpn}(R_4)^{iq}_r-4(\Lambda_U)^{jml}(\Lambda_D)_{lkn}(\Lambda_U)_{pqm}(Y_d)^{np}(R_7)^{iq}
\nonumber\\&&-4(\Lambda_U)^{jml}(\Lambda_D)_{lkn}(\Lambda_U)_{qpm}(\Lambda_D)^{prn}(R_1)^{iq}_r-4(\Lambda_U)^{jml}(\Lambda_D)_{lkn}(\Lambda_U)_{qpm}(Y_d)^{np}(R_9)^{iq}
\nonumber\\&&-4(R_1)^{ij}_k[C_L]^2-2[C_Q]^2(R_1)^{ji}_k-2[C_{u^c}]^2(R_1)^{ij}_k
 \nonumber\\&&-4(R_1)^{ij}_kC_{u^c}C_L-4(R_1)^{ji}_kC_QC_L
             \nonumber\\&&-2(R_4)^{il}_m(\Lambda_D)_{nkl}(\Lambda_U)^{jnm}C_{u^c}+2(R_7)^{il}(\Lambda_D)_{lkn}(Y_u)^{nj}C_{u^c}
\nonumber\\&&-2(R_3)^{jl}_m(\Lambda_E)_{lnk}(\Lambda_D)^{mni}C_Q-2(R_3)^{jl}_m(Y_e)_{kl}(Y_d)^{im}C_Q
\nonumber\\&&-2(R_9)^{lj}(\Lambda_D)_{nkl}(Y_d)^{in}C_Q+2(R_1)^{lj}_m(\Lambda_D)_{nkl}(\Lambda_D)^{nmi}C_Q
\nonumber\\&&+2(R_4)^{il}_m(\Lambda_D)_{nkl}(\Lambda_U)^{jnm}C_L+2(R_4)^{il}_m(\Lambda_D)_{lkn}(Y_u)^{jn}C_L
\nonumber\\&&+2(R_1)^{lj}_m(\Lambda_D)_{nkl}(\Lambda_D)^{nmi}C_L-2(R_9)^{lj}(\Lambda_D)_{nkl}(Y_d)^{in}C_L
\nonumber\\&&-(R_3)^{jl}_m(\Lambda_E)_{lkn}(\Lambda_D)^{mni}C_L-2(R_3)^{jl}_m(Y_e)_{kl}(Y_d)^{im}C_L
\nonumber\\&&-12(Y_u)^{ij}(Y_u)_{mn}(\Lambda_U)^{nrq}(\Lambda_D)_{rkp}(R_4)^{mp}_q+4(Y_u)^{ij}(Y_u)_{mn}(Y_u)^{rn}(\Lambda_D)_{pkr}(R_7)^{mp}
\nonumber\\&&+12(Y_u)^{ij}(Y_u)_{nm}(\Lambda_D)^{rqn}(\Lambda_D)_{rkp}(R_1)^{pm}_q-12(Y_u)^{ij}(Y_u)_{nm}(Y_u)^{nr}(\Lambda_D)_{rkp}(R_9)^{pm}
\nonumber\\&&-12(Y_u)^{ij}(Y_u)_{nm}(\Lambda_D)^{qrn}(\Lambda_E)_{prk}(R_3)^{mp}_q-12(Y_u)^{ij}(Y_u)_{nm}(Y_d)^{nq}(Y_e)_{kp}(R_3)^{mp}_q
\nonumber\\&&+(\Lambda_D)^{lmi}(\Lambda_D)_{lkn}(Y_u)^{nj}(\Lambda_D)_{qmr}(R_7)^{rq}-(Y_d)^{il}(\Lambda_D)_{lkn}(Y_u)^{nj}(Y_d)_{rq}(R_7)^{rq}
\nonumber\\&&-(\Lambda_D)^{lmi}(\Lambda_D)_{lkn}(Y_u)^{nj}(\Lambda_E)_{rqm}(R_5)^{qr}-(Y_d)^{il}(\Lambda_D)_{lkn}(Y_u)^{nj}(Y_e)_{qr}(R_5)^{qr}
\nonumber\\&&-(\Lambda_D)^{lmi}(\Lambda_D)_{lkn}(Y_u)^{nj}(Y_e)_{mq}(R_2)^{q}-(Y_u)^{lj}(\Lambda_D)_{nkl}(\Lambda_D)^{npi}(Y_u)_{qr}(R_1)^{qr}_p
\nonumber\\&&-3(Y_u)^{lj}(\Lambda_D)_{nkl}(Y_d)^{in}(Y_u)_{qr}(R_9)^{qr}-2(Y_u)^{ip}(Y_u)_{np}(Y_u)^{lj}(\Lambda_D)_{rkl}(R_7)^{nr}
\nonumber\\&&-4(Y_u)^{ip}(\Lambda_U)_{pnm}(\Lambda_U)^{jml}(\Lambda_D)_{lkr}(R_7)^{nr}+4(\Lambda_D)^{qpi}(\Lambda_D)_{mpn}(\Lambda_U)^{jml}(\Lambda_D)_{lkr}(R_4)^{nr}_q
\nonumber\\&&+4(Y_d)^{iq}(Y_d)_{nm}(\Lambda_U)^{jml}(\Lambda_D)_{lkr}(R_4)^{nr}_q+2(\Lambda_D)^{pqi}(\Lambda_U)_{npm}(\Lambda_U)^{jml}(\Lambda_D)_{lkr}(R_1)^{rn}_q
\nonumber\\&&-4(Y_d)^{ip}(\Lambda_U)_{npm}(\Lambda_U)^{jml}(\Lambda_D)_{lkr}(R_9)^{rn}+4(Y_u)^{pj}(\Lambda_D)_{mnp}(\Lambda_D)^{mli}(\Lambda_E)_{rlk}(R_5)^{nr}
\nonumber\\&&-4(Y_u)^{pj}(Y_d)_{pm}(\Lambda_D)^{mli}(\Lambda_E)_{rlk}(R_2)^{r}-4(Y_u)^{pj}(Y_d)_{pm}(Y_d)^{im}(Y_e)_{kr}(R_2)^{r}
\nonumber\\&&+4(Y_u)^{pj}(\Lambda_D)_{mnp}(Y_d)^{im}(Y_e)_{kr}(R_5)^{nr}-4(\Lambda_U)^{jpq}(\Lambda_D)_{pmn}(\Lambda_D)^{lmi}(\Lambda_D)_{lkr}(R_4)^{nr}_q
\nonumber\\&&+4(\Lambda_U)^{jpq}(\Lambda_D)_{pmn}(Y_d)^{il}(\Lambda_D)_{lkr}(R_4)^{nr}_q-18(Y_u)^{ij}(Y_u)_{lm}(R_1)^{lm}_kC_L
\nonumber\\&&-(\Lambda_D)^{mli}(\Lambda_E)_{nkl}(R_3)^{jn}_mC_{u^c}+2(\Lambda_D)^{lmi}(\Lambda_D)_{lkn}(R_1)^{nj}_mC_{u^c}
\nonumber\\&&-2(Y_d)^{il}(\Lambda_D)_{lkn}(R_9)^{nj}C_{u^c}-4(\Lambda_U)^{jml}(\Lambda_D)_{lkn}(R_4)^{ni}_mC_Q
\nonumber\\&&+2(Y_u)^{lj}(\Lambda_D)_{nkl}(R_7)^{in}C_Q+4(R_4)^{li}_nC_Q(\Lambda_U)^{jqn}(\Lambda_D)_{qkl}
\nonumber\\&&-2(R_7)^{il}C_Q(Y_u)^{qj}(\Lambda_D)_{lkq}-2(R_1)^{lj}_nC_{u^c}(\Lambda_D)^{qni}(\Lambda_D)_{qkl}
\nonumber\\&&+2(R_9)^{lj}C_{u^c}(Y_d)^{iq}(\Lambda_D)_{qkl}+2(R_3)^{jl}_nC_{u^c}(\Lambda_D)^{nqi}(\Lambda_E)_{lqk}
\nonumber\\&&+2(R_3)^{jl}_nC_{u^c}(Y_d)^{in}(Y_e)_{kl}+4(R_1)^{ij}_k[C_{u^c}]2
\nonumber\\&&+4(R_1)^{ij}_k[C_Q]^2-2(R_1)^{il}_k(\Lambda_U)_{lmn}(\Lambda_U)^{jmp}(\gamma_D)^n_p
\nonumber\\&&-2(R_1)^{il}_k(Y_u)_{nl}(Y_u)^{pj}(\gamma_Q)^n_p-2(R_1)^{il}_k(Y_u)_{ml}(Y_u)^{mj}(\gamma_{H_2})
\nonumber\\&&+(R_1)^{lj}_k(\Lambda_D)_{mnl}(\Lambda_D)^{mpi}(\gamma_L)^n_p-(R_1)^{lj}_k(\Lambda_D)_{nml}(\Lambda_D)^{pmi}(\gamma_D)^n_p
\nonumber\\&&-(R_1)^{lj}_k(Y_u)_{lk}(Y_u)^{ip}(\gamma_U)^n_p-(R_1)^{lj}_k(Y_u)_{lm}(Y_u)^{im}(\gamma_{H_2})
\nonumber\\&&-(R_1)^{lj}_k(Y_d)_{lm}(Y_d)^{im}(\gamma_{H_1})-(R_1)^{lj}_k(Y_d)_{ln}(Y_d)^{ip}(\gamma_D)^n_p
\nonumber\\&&+(R_1)^{lj}_k(\Lambda_D)_{mnl}(Y_d)^{im}(\gamma_{H_1L})_n+(R_1)^{lj}_k(Y_d)_{lm}(\Lambda_D)^{mpi}(\gamma_{H_1L})^p
\nonumber\\&&-6(Y_u)_{lm}(Y_u)^{ij}(R_1)^{mp}_k(\gamma_Q)^l_p-6(Y_u)_{ml}(Y_u)^{ij}(R_1)^{pm}_k(\gamma_Q)^l_p
\nonumber\\&&-2(R_1)^{il}_kC_{u^c}(\gamma_U)^j_l-2(R_1)^{lj}_kC_Q(\gamma_Q)^i_l-2(R_1)^{ij}_lC_L(\gamma_L)^l_k
\nonumber\\&&+2(\Lambda_D)^{lmi}(\Lambda_D)_{nkp}(\gamma_D)^n_l(R_1)^{pj}_m+2(Y_u)^{il}(\Lambda_D)_{nkp}(\gamma_D)^n_l(R_9)^{pj}
\nonumber\\&&-2(\Lambda_D)^{mli}(\Lambda_E)_{pnk}(\gamma_L)^n_l(R_3)^{jp}_m-2(Y_d)^{im}(Y_e)_{kp}(\gamma_{H_1})(R_3)^{jp}_m
\nonumber\\&&+4(\Lambda_U)^{jml}(\Lambda_D)_{nkp}(\gamma_D)^n_l(R_4)^{ip}_m+2(Y_u)^{lj}(\Lambda_D)_{pkn}(\gamma_Q)^n_l(R_7)^{ip}
\nonumber\\&&+2(Y_d)^{im}(\Lambda_E)_{pnk}(\gamma_{H_1L})^n(R_3)^{jp}_m+2(\Lambda_D)^{mli}(Y_e)_{kp}(\gamma_{H_1L})_l(R_3)^{jp}_m
\nonumber\\&&-2(\Lambda_D)^{mli}(\Lambda_E)_{nlk}(R_3)^{nj}_p (\gamma_D)^p_m-2(Y_d)^{im}(Y_e)_{kn}(R_3)^{jn}_p (\gamma_D)^p_m
\nonumber\\&&+2(\Lambda_D)^{lmi}(\Lambda_D)_{lkn}(R_1)^{nj}_p (\gamma_L)^p_m-2(Y_d)^{il}(\Lambda_D)_{lkn}(R_9)^{nj} (\gamma_{H_1})
\nonumber\\&&+4(\Lambda_U)^{jml}(\Lambda_D)_{lkn}(R_4)^{ni}_p (\gamma_D)^p_m+2(Y_u)^{lj}(\Lambda_D)_{nkl}(R_7)^{in} (\gamma_{H_2})
\nonumber\\&&-2(Y_d)^{il}(\Lambda_D)_{lkn}(R_1)^{nj}_p (\gamma_{H_1L})^p +2(\Lambda_D)^{lmi}(\Lambda_D)_{lkn}(R_9)^{nj} (\gamma_{H_1L})_m
\nonumber\\&&+2(\Lambda_D)^{lmi}(\Lambda_D)_{lkn}(\gamma_Q)^n_p(R_1)^{jp}_m-2(Y_d)^{il}(\Lambda_D)_{lkn}(\gamma_Q)^n_p(R_9)^{pj}
\nonumber\\&&+2(\Lambda_D)^{mli}(Y_e)_{kn}(\gamma_E)^n_p(R_3)^{jp}_m-2(Y_d)^{im}(Y_e)_{kn}(\gamma_E)^n_p(R_3)^{jp}_m
\nonumber\\&&+4(\Lambda_U)^{jml}(\Lambda_D)_{lkn}(\gamma_Q)^n_p(R_4)^{ip}_m+2(Y_u)^{lj}(\Lambda_D)_{nkl}(\gamma_D)^n_p(R_7)^{ip}
\nonumber\\&&-3(R_1)^{ij}_l(\Lambda_D)^{nlm}(\Lambda_D)_{pkm}(\gamma_D)^p_n-3(R_1)^{ij}_l(\Lambda_D)^{mln}(\Lambda_D)_{mkp}(\gamma_Q)^p_n
\nonumber\\&&+3(R_9)^{ij}(Y_d)^{mn}(\Lambda_D)_{pkm}(\gamma_D)^p_n+3(R_9)^{ij}(Y_d)^{nm}(\Lambda_D)_{mkp}(\gamma_Q)^p_n
\nonumber\\&&-2(R_1)^{ij}_l(\Lambda_E)^{mln}(\Lambda_E)_{mkp}(\gamma_L)^p_n-(R_1)^{ij}_l(\Lambda_E)^{nlm}(\Lambda_E)_{pkm}(\gamma_E)^p_n
\nonumber\\&&-(R_1)^{ij}_l(Y_e)^{ln}(Y_e)_{kp}(\gamma_E)^p_n-(R_1)^{ij}_l(Y_e)^{lm}(Y_e)_{km}(\gamma_{H_1})
\nonumber\\&&+(R_1)^{ij}_l(Y_e)^{lm}(\Lambda_E)_{mpk}(\gamma_{LH_1})^p+(R_1)^{ij}_l(\Lambda_E)^{mln}(Y_e)_{km}(\gamma_{LH_1})_n
\nonumber\\&&-2(R_1)^{ij}_k[\frac{99}{100}g^4_1+\frac{18}{4}g^4_2]+2(R_1)^{ij}_k[\frac{132}{75}g^4_1-4g^4_3]
\nonumber\\&&+2(R_1)^{ij}_k[\frac{33}{300}g^4_1-\frac{3}{4}g^4_2-4g^4_3]
\end{eqnarray}

\begin{eqnarray}
(16\pi^2)^2(\beta^{(2)}_{R_2})^i&=&
-6(Y_d)^{pl}(Y_u)_{pn}C_{u^c}(R_3)^{ni}_l-6(Y_d)^{pn}(Y_u)_{pm}(R_3)^{mi}_nC_{d_c}
\nonumber\\&&-12(Y_d)^{lm}(Y_u)_{ln}(\Lambda_U)_{qpm}(\Lambda_U)^{nrp}(R_3)^{iq}_r-12(Y_d)^{lm}(Y_u)_{ln}(\Lambda_D)_{mpq}(Y_u)^{pn}(R_5)^{qi}
\nonumber\\&&-12(Y_d)^{lm}(Y_u)_{ln}(Y_d)_{pm}(Y_u)^{pn}(R_2)^{i}-4(R_2)^{i}[C_{H_2}]^2-6(R_3)^{li}_m(Y_u)_{nl}(Y_d)^{nm}C_{H_1}
\nonumber\\&&-2[C_{H_1}]^2(R_2)^{i}-2[C_{e^c}]^2(R_2)^{i}-4(R_2)^{i}C_{H_2}C_{H_1}
\nonumber\\&&-4(R_2)^iC_{e^c}C_{H_2}-6(R_3)^{li}_m(Y_u)_{nl}(Y_d)^{nm}C_{H_2}
\nonumber\\&&-6(Y_e)^{li}(\Lambda_D)_{nlm}(Y_d)^{rn}(Y_u)_{rp}(R_9)^{mp}+6(Y_e)^{li}(\Lambda_D)_{nlm}(\Lambda_D)^{nqr}(Y_u)_{rp}(R_1)^{mp}_q
\nonumber\\&&-6(Y_e)^{li}(\Lambda_D)_{nlm}(\Lambda_U)^{rnq}(Y_u)_{pr}(R_4)^{mp}_q+6(Y_e)^{li}(\Lambda_D)_{nlm}(Y_u)^{nr}(Y_u)_{pr}(R_7)^{pm}
\nonumber\\&&-12(Y_e)^{pi}(\Lambda_D)_{mpn}(Y_d)^{lm}(Y_u)_{lr}(R_9)^{nr}-12(\Lambda_E)^{iqp}(\Lambda_D)_{mpn}(Y_d)^{lm}(Y_u)_{lr}(R_1)^{nr}_q
\nonumber\\&&-18(Y_e)^{ni}(\Lambda_D)_{lnm}(R_7)^{ml}C_{H_2}-18(Y_e)^{ni}(\Lambda_E)_{lnm}(R_5)^{ml}C_{H_2}
\nonumber\\&&-6(Y_e)^{ni}(Y_e)_{nl}(R_2)^{l}C_{H_2}-6(Y_d)^{lm}(Y_d)_{ln}(R_3)^{ni}_mC_{e^c}
\nonumber\\&&+6(R_3)^{li}_nC_{e^c}(Y_d)^{qn}(Y_u)_{ql}+4(R_2)^{i}[C_{e^c}]^2+4(R_2)^{i}[C_{H_1}]^2
\nonumber\\&&-2(R_2)^{l}(\Lambda_E)_{lnm}(\Lambda_E)^{ipm}(\gamma_L)^n_p-2(R_2)^{l}(Y_e)_{ml}(Y_e)^{mi}(\gamma_{H_1})
\nonumber\\&&-2(R_2)^{l}(Y_e)_{nl}(Y_e)^{pi}(\gamma_L)^n_p-(R_2)^{i}(Y_d)_{mn}(Y_d)^{mp}(\gamma_D)^n_p
\nonumber\\&&+3(R_2)^{i}(Y_d)_{nm}(Y_d)^{pm}(\gamma_Q)^n_p-(R_2)^{i}(Y_e)_{nm}(Y_e)^{pm}(\gamma_L)^n_p
\nonumber\\&&-(R_5)^{li}(\Lambda_E)_{mnl}(Y_e)^{pm}(\gamma_L)^n_p-(R_5)^{il}(\Lambda_E)_{nml}(Y_e)^{mp}(\gamma_E)^n_p
\nonumber\\&&+3(R_5)^{li}(\Lambda_D)_{nlm}(Y_e)^{mp}(\gamma_D)^n_p+3(R_5)^{li}(\Lambda_D)_{mln}(Y_d)^{pm}(\gamma_Q)^n_p
\nonumber\\&&+2(R_2)^{l}(Y_e)_{ml}(\Lambda_E)^{ipm}(\gamma_{LH_1})_p-(R_2)^{l}(\Lambda_E)_{lmn}(Y_e)^{mi}(\gamma_{LH_1})^n
\nonumber\\&&+(R_5)^{li}(Y_e)_{lm}(Y_e)^{pm}(\gamma_{LH_1})_p-2(Y_e)_{nl}(Y_e)^{ni}(R_2)^{p}(\gamma_E)^l_p-
\nonumber\\&&-2(Y_e)_{nm}(Y_e)^{ni}(R_2)^{m}(\gamma_{H_1})-2(\Lambda_E)_{lnm}(Y_e)^{ni}(R_5)^{mp}(\gamma_E)^l_p
\nonumber\\&&-2(\Lambda_E)_{mnl}(Y_e)^{in}(R_5)^{pm}(\gamma_L)^l_p-6(\Lambda_D)_{lnm}(Y_e)^{ni}(R_7)^{mp}(\gamma_D)^l_p
\nonumber\\&&-6(\Lambda_D)_{mnl}(Y_e)^{ni}(R_7)^{pm}(\gamma_Q)^l_p-2(\Lambda_E)_{mnl}(Y_e)^{in}(R_2)^{m}(\gamma_{LH_1})^l
\nonumber\\&&-2(Y_e)_{mn}(Y_e)^{ni}(R_5)^{pm}(\gamma_{LH_1})_p-2(R_2)^{l}C_{e^c}(\gamma_E)^i_l-2(R_2)^{i}[C_{H_1}\gamma_{H_1}]
\nonumber\\&&-2(R_2)^{i}[C_{H_2}\gamma_{H_2}]-6(Y_d)^{lm}(Y_u)_{np}(\gamma_Q)^n_l(R_3)^{ip}_m
\nonumber\\&&-6(Y_d)^{lm}(Y_u)_{ln}(R_3)^{ni}_p(\gamma_D)^p_m-6(Y_d)^{lm}(Y_u)_{ln}(\gamma_U)^n_p(R_3)^{pi}_m
\nonumber\\&&-3(R_2)^{i}(Y_u)^{mn}(Y_u)_{mp}(\gamma_U)^p_n-3(R_2)^{i}(Y_u)^{nm}(Y_u)_{pm}(\gamma_Q)^p_n
\nonumber\\&&-2(R_2)^{i}[\frac{99}{100}g^4_1+\frac{3}{4}g^4_2]+2(R_2)^{l}[\frac{99}{15}g^4_1]
\nonumber\\&&+2(R_2)^{i}[\frac{99}{100}+\frac{3}{4}]
\end{eqnarray}

\begin{eqnarray}
(16\pi^2)^2(\beta^{(2)}_{R_3})^{ij}_k&=&
-4(Y_d)^{jp}(\Lambda_D)_{kpn}C_Q(R_9)^{ni}+4(\Lambda_E)^{jlp}(\Lambda_D)_{kpn}C_Q(R_1)^{ni}_l
\nonumber\\&&-4(Y_e)^{lj}(Y_d)_{nk}C_Q(R_1)^{ni}_l+4(Y_u)^{pi}(\Lambda_D)_{kpn}C_L(R_5)^{nj}
\nonumber\\&&-4(\Lambda_U)^{ilp}(\Lambda_U)_{npk}C_{u^c}(R_3)^{nj}_l-4(Y_e)^{pj}(\Lambda_D)_{kpm}(R_9)^{mi}C_{H_1}
\nonumber\\&&+4(\Lambda_E)^{jnp}(\Lambda_D)_{kpm}(R_1)^{mi}_nC_L-4(Y_e)^{nj}(Y_d)_{mk}(R_1)^{mi}_nC_L
\nonumber\\&&-4(Y_u)^{pi}(Y_d)_{pk}(R_2)^{j}C_{H_2}+4(Y_u)^{pi}(\Lambda_D)_{kpm}(R_5)^{mj}C_{H_2}
\nonumber\\&&+4(Y_e)^{mj}(Y_d)_{nk}(\Lambda_E)_{qpm}(\Lambda_D)^{rpn}(R_3)^{iq}_r-4(Y_e)^{mj}(Y_d)_{nk}(Y_e)_{mq}(Y_d)^{rn}(R_3)^{iq}_r
\nonumber\\&&-4(Y_e)^{lj}(\Lambda_D)_{kln}(Y_e)_{pq}(\Lambda_D)^{rpn}(R_3)^{iq}_r+4(\Lambda_E)^{jml}(\Lambda_D)_{kln}(\Lambda_E)_{qpm}(\Lambda_D)^{rpn}(R_3)^{iq}_r
\nonumber\\&&+4(\Lambda_E)^{jml}(\Lambda_D)_{kln}(Y_e)_{mq}(Y_d)^{nr}(R_3)^{iq}_r+4(Y_e)^{mj}(Y_d)_{nk}(\Lambda_D)_{pmq}(\Lambda_D)^{prn}(R_1)^{qi}_r
\nonumber\\&&-4(\Lambda_E)^{jml}(\Lambda_D)_{kln}(\Lambda_D)_{pmq}(\Lambda_D)^{prn}(R_1)^{qi}_r-4(Y_e)^{lj}(\Lambda_D)_{kln}(Y_d)_{qp}(\Lambda_D)^{prn}(R_1)^{qi}_r
\nonumber\\&&-4(Y_e)^{lj}(\Lambda_D)_{kln}(Y_d)_{qp}(Y_d)^{np}(R_9)^{qi}+4(\Lambda_E)^{jml}(\Lambda_D)_{kln}(\Lambda_D)_{pmq}(Y_d)^{np}(R_9)^{qi}
\nonumber\\&&-8(\Lambda_U)^{iml}(\Lambda_U)_{nlk}(\Lambda_U)_{qpm}(\Lambda_U)^{nrp}(R_3)^{qj}_r-12(Y_u)^{li}(\Lambda_D)_{knl}(Y_u)_{pq}(\Lambda_D)^{rnp}(R_3)^{qi}_r
\nonumber\\&&-12(Y_u)^{li}(Y_d)_{lk}(Y_u)_{pq}(Y_d)^{pr}(R_3)^{qj}_r-4(Y_e)^{mj}(Y_d)_{nk}(\Lambda_D)_{pmq}(Y_d)^{np}(R_9)^{qi}
\nonumber\\&&-4(R_3)^{ij}_k[C_{d_c}]^2-2[C_{e^c}]^2(R_3)^{ij}_k-2[C_{u^c}]^2(R_3)^{ij}_k
\nonumber\\&&-4(R_3)^{ij}_kC_{u^c}C_{d^c}-4(R_3)^{ij}_kC_{e^c}C_{d_c}
\nonumber\\&&+2(R_5)^{lj}(\Lambda_U)_{kln}(Y_u)^{ni}C_{u^c}-4(R_2)^{j}(Y_d)_{nk}(Y_u)^{ni}C_{u^c}
\nonumber\\&&+4(R_3)^{lj}_m(\Lambda_U)_{lnk}(\Lambda_U)^{inm}C_{u^c}-4(R_9)^{li}(\Lambda_D)_{knl}(Y_u)^{nj}C_{e^c}
\nonumber\\&&-4(R_1)^{li}_m(\Lambda_D)_{knl}(\Lambda_E)^{jnm}C_{e^c}-4(R_2)^{j}(Y_d)_{nk}(Y_u)^{ni}C_{d_c}
\nonumber\\&&+4(R_5)^{lj}(\Lambda_D)_{kln}(Y_u)^{ni}C_{d_c}+4(R_3)^{lj}_m(\Lambda_U)_{lkr}(\Lambda_U)^{inm}C_{d_c}
\nonumber\\&&-4(R_1)^{li}_m(\Lambda_D)_{knl}(\Lambda_E)^{jnm}C_{d_c}-4(R_9)^{li}(\Lambda_D)_{knl}(Y_e)^{nj}C_{d_c}
\nonumber\\&&+2(\Lambda_E)^{jml}(\Lambda_D)_{kln}(Y_u)^{ni}(Y_e)_{mr}(R_2)^{r}-2(Y_e)^{jm}(Y_d)_{nk}(Y_u)^{ni}(Y_e)_{mr}(R_2)^{r}
\nonumber\\&&-2(Y_e)^{mj}(Y_d)_{nk}(Y_u)^{ni}(\Lambda_E)_{rmq}(R_5)^{qr}-2(\Lambda_E)^{jml}(\Lambda_D)_{kln}(Y_u)^{ni}(\Lambda_E)_{rqm}(R_5)^{qr}
\nonumber\\&&+2(Y_e)^{lj}(\Lambda_D)_{kln}(Y_u)^{ni}(Y_e)_{qr}(R_5)^{qr}+6(Y_u)^{li}(\Lambda_D)_{knl}(\Lambda_E)^{jpn}(Y_u)_{qr}(R_1)^{qr}_p
\nonumber\\&&-6(Y_u)^{li}(Y_d)_{lk}(Y_e)^{pj}(Y_u)_{qr}(R_1)^{qr}_p-6(Y_u)^{li}(\Lambda_D)_{knl}(Y_e)^{nj}(Y_u)_{qr}(R_9)^{qr}
\nonumber\\&&-4(Y_u)^{qi}(\Lambda_D)_{nmq}(\Lambda_E)^{jml}(\Lambda_D)_{klr}(R_7)^{rn}+4(Y_u)^{qi}(\Lambda_D)_{nmq}(Y_e)^{mj}(Y_d)_{rk}(R_7)^{rn}
\nonumber\\&&-4(Y_u)^{qi}(Y_d)_{qn}(Y_e)^{jl}(\Lambda_D)_{klr}(R_7)^{rn}+4(\Lambda_U)^{iqp}(\Lambda_D)_{qmn}
(\Lambda_E)^{jml}(\Lambda_D)_{klr}(R_4)^{nr}_p
\nonumber\\&&+8(\Lambda_U)^{iqp}(\Lambda_D)_{qmn}(Y_e)^{mj}(Y_d)_{rk}(R_4)^{nr}_p+8(\Lambda_U)^{iqp}(Y_d)_{qn}(Y_e)^{lj}(\Lambda_D)_{klr}(R_4)^{nr}_p
\nonumber\\&&+4(\Lambda_E)^{jlm}(\Lambda_D)_{kln}(R_1)^{ni}_mC_{u^c}+4(Y_e)^{mj}(Y_d)_{nk}(R_1)^{ni}_mC_{u^c}
\nonumber\\&&-4(Y_e)^{lj}(\Lambda_D)_{kln}(R_9)^{ni}C_{u^c}+4(\Lambda_U)^{iml}(\Lambda_D)_{nlk}(R_3)^{nj}_lC_{e^c}
\nonumber\\&&-4(Y_u)^{li}(Y_d)_{lk}(R_2)^{j}C_{e^c}+4(Y_u)^{li}(\Lambda_D)_{knl}(R_5)^{nj}C_{e^c}
\nonumber\\&&-4(R_3)^{lj}_nC_{e^c}(\Lambda_U)^{iqn}(\Lambda_U)_{lqk}+4(R_2)^{j}C_{e^c}(Y_u)^{qi}(Y_d)_{qk}
\nonumber\\&&-4(R_2)^{j}C_{e^c}(Y_u)^{qi}(\Lambda_D)_{klq}+4(R_1)^{li}_nC_{u^c}(\Lambda_E)^{jqn}(\Lambda_D)_{kql}
\nonumber\\&&+4(R_9)^{li}C_{u^c}(Y_e)^{qj}(\Lambda_D)_{kql}+4(R_3)^{ij}_k[C_{u^c}]^2+4(R_3)^{ij}_k[C_{e^c}]^2
\nonumber\\&&+2(R_3)^{lj}_k(\Lambda_U)_{lnm}(\Lambda_U)^{ipm}(\gamma_D)^n_p-2(R_3)^{lj}_k(Y_u)_{nl}(Y_u)^{pi}(\gamma_Q)^n_p
\nonumber\\&&-2(R_3)^{lj}_k(Y_u)_{ml}(Y_u)^{mi}(\gamma_{H_2})-2(R_3)^{lj}_kC_{u^c}(\gamma_U)^i_l-2(R_3)^{il}_kC_{e^c}(\gamma_E)^j_l
\nonumber\\&&-2(R_3)^{ij}_kC_{d_c}(\gamma_D)^l_k+4(\Lambda_E)^{jml}(\Lambda_D)_{knp}(\gamma_L)^n_l(R_1)^{pi}_m
\nonumber\\&&-4(Y_e)^{mj}(Y_d)_{pk}(\gamma_{H_1})(R_1)^{pi}_m-4(Y_e)^{lj}(\Lambda_D)_{knp}(\gamma_L)^n_l(R_9)^{pi}
\nonumber\\&&-4(\Lambda_U)^{iml}(\Lambda_U)_{pnk}(\gamma_D)^n_l(R_3)^{pj}_m-4(Y_u)^{li}(Y_d)_{nk}(\gamma_Q)^l_n(R_2)^{j}
\nonumber\\&&+4(\Lambda_E)^{jml}(\Lambda_D)_{kln}(R_1)^{ni}_p(\gamma_L)^p_m-4(Y_e)^{mj}(Y_d)_{nk}(R_1)^{ni}_p(\gamma_L)^p_m
\nonumber\\&&-4(Y_e)^{lj}(\Lambda_D)_{kln}(R_9)^{ni}(\gamma_{H_1})-4(\Lambda_U)^{iml}(\Lambda_U)_{nlk}(R_3)^{nj}_p(\gamma_D)^p_m
\nonumber\\&&+4(\Lambda_E)^{jlm}(\Lambda_D)_{kln}(R_9)^{ni}(\gamma_{LH_1})_m-4(Y_e)^{lj}(\Lambda_D)_{kln}(R_1)^{ni}_p(\gamma_{LH_1})^p
\nonumber\\&&-4(Y_u)^{li}(Y_d)_{lk}(R_2)^{j}(\gamma_{H_2})+4(Y_u)^{li}(\Lambda_D)_{knl}(R_2)^{j}(\gamma_{H_2})
\nonumber\\&&+4(\Lambda_E)^{jml}(\Lambda_D)_{kln}(\gamma_Q)^n_p(R_1)^{pi}_m-4(Y_E)^{mj}(Y_d)_{nk}(\gamma_Q)^n_p(R_1)^{pi}_m
\nonumber\\&&-4(Y_e)^{jl}(\Lambda_D)_{kln}(\gamma_Q)^n_p(R_9)^{pi}+4(Y_e)^{mj}(\Lambda_D)_{kln}(\gamma_Q)^n_p(R_1)^{pi}_m
\nonumber\\&&+4(\Lambda_U)^{iml}(\Lambda_U)_{nlk}(\gamma_U)^n_p(R_3)^{pj}_m-4(Y_u)^{li}(Y_d)_{lk}(\gamma_{H_1})(R_2)^{j}
\nonumber\\&&+4(Y_u)^{li}(\Lambda_D)_{knl}(\gamma_L)^n_p(R_5)^{pj}+4(Y_u)^{li}(\Lambda_D)_{knl}(\gamma_{LH_1})^n(R_2)^{j}
\nonumber\\&&+4(Y_u)^{li}(Y_d)_{lk}(\gamma_{LH_1})_p(R_5)^{pj}-2(R_3)^{ij}_l(\Lambda_D)^{lmn}(\Lambda_D)_{kmp}(\gamma_Q)^p_n
\nonumber\\&&+2(R_3)^{ij}_l(\Lambda_D)^{lnm}(\Lambda_D)_{kpm}(\gamma_L)^p_n-2(R_3)^{ij}_l(Y_d)^{nl}(Y_d)_{pk}(\gamma_Q)^p_n
\nonumber\\&&-2(R_3)^{ij}_l(Y_d)^{ml}(Y_d)_{mk}(\gamma_{H_1})-2(R_3)^{ij}_l(\Lambda_U)^{nml}(\Lambda_U)_{pmk}(\gamma_U)^p_n
\nonumber\\&&-2(R_3)^{ij}_l(\Lambda_U)^{mnl}(\Lambda_U)_{mpk}(\gamma_D)^p_n+2(R_3)^{ij}_l(Y_e)^{ml}(\Lambda_D)^{kpm}(\gamma_{LH_1})^p
\nonumber\\&&+2(R_3)^{ij}_l(\Lambda_D)^{lnm}(Y_d)_{mk}(\gamma_{LH_1})_n+2(R_3)^{ij}_k[\frac{99}{25}g^4_1]
\nonumber\\&&-2(R_3)^{ij}_k[\frac{33}{75}g^4_1-4g^4_3]+2(R_3)^{ij}_k[\frac{132}{75}g^4_1-4g^4_3]
\end{eqnarray}

\begin{eqnarray}
(16\pi^2)^2(\beta^{(2)}_{R_4})^{ij}_k&=&
-2(Y_u)^{jp}(\Lambda_U)_{pnk}C_{d_c}(R_7)^{in}+2(Y_d)^{jp}(\Lambda_U)_{npk}C_{d_c}(R_7)^{in}
\nonumber\\&&-2(\Lambda_D)^{ljp}(\Lambda_U)_{npk}C_{d_c}(R_1)^{in}_l-2(\Lambda_D)^{lpj}(\Lambda_D)_{kpn}C_Q(R_4)^{in}_l
\nonumber\\&&-2(Y_d)^{jl}(Y_d)_{nk}C_Q(R_4)^{in}_l-2(Y_u)^{jp}(\Lambda_U)_{pmk}(R_7)^{im}C_{H_2}
\nonumber\\&&+2(Y_d)^{jp}(\Lambda_U)_{mpk}(R_9)^{im}C_{H_1}-2(\Lambda_D)^{plj}(\Lambda_U)_{mpk}(R_1)^{im}_lC_L
\nonumber\\&&+2(\Lambda_D)^{npj}(\Lambda_D)_{kpm}(R_4)^{im}_nC_{d_c}+2(Y_d)^{jn}(Y_d)_{mk}(R_4)^{im}_nC_{d_c}
\nonumber\\&&+4(Y_d)^{jl}(\Lambda_U)_{nlk}(Y_d)_{qp}(\Lambda_U)^{nrp}(R_4)^{iq}_r+4(\Lambda_D)^{lmj}(\Lambda_U)_{nlk}(\Lambda_D)_{pmq}(\Lambda_U)^{nrp}(R_4)^{iq}_r
\nonumber\\&&-4(Y_u)^{jl}(\Lambda_U)_{lnk}(Y_u)_{qp}(\Lambda_U)^{prn}(R_4)^{iq}_r-2(\Lambda_D)^{mlj}(\Lambda_D)_{kln}(\Lambda_D)_{mpq}(\Lambda_D)^{rpn}(R_4)^{iq}_r
\nonumber\\&&-2(\Lambda_D)^{mlj}(\Lambda_D)_{kln}(Y_d)_{qm}(Y_d)^{nr}(R_4)^{iq}_r-2(Y_d)^{jm}(Y_d)_{nk}(\Lambda_D)_{mpq}(\Lambda_D)^{rpn}(R_4)^{iq}_r
\nonumber\\&&-2(Y_d)^{jm}(Y_d)_{nk}(Y_d)_{qm}(Y_d)^{nr}(R_4)^{iq}_r-2(Y_u)^{jl}(\Lambda_U)_{lnk}(Y_u)_{pq}(Y_d)^{pn}(R_9)^{iq}
\nonumber\\&&+2(Y_u)^{jl}(\Lambda_U)_{lnk}(Y_u)_{pq}(\Lambda_D)^{nrp}(R_1)^{iq}_r-2(\Lambda_D)^{mlj}(\Lambda_D)_{kln}(\Lambda_U)_{qpm}(Y_d)^{np}(R_9)^{iq}
\nonumber\\&&-2(Y_d)^{jm}(Y_d)_{nk}(\Lambda_U)_{qpm}(Y_d)^{np}(R_9)^{iq}+2(\Lambda_D)^{mlj}(\Lambda_D)_{kln}(\Lambda_U)_{qpm}(\Lambda_D)^{prn}(R_1)^{iq}_r
\nonumber\\&&+2(Y_d)^{jm}(Y_d)_{nk}(\Lambda_U)_{qpm}(\Lambda_D)^{prn}(R_1)^{iq}_r+2(Y_d)^{jm}(Y_d)_{nk}(\Lambda_U)_{pmq}(Y_u)^{np}(R_7)^{iq}
\nonumber\\&&+2(\Lambda_D)^{mlj}(\Lambda_D)_{kln}(\Lambda_U)_{pqm}(Y_u)^{np}(R_7)^{iq}-2(R_4)^{ji}_k[C_{d_c}]^2
\nonumber\\&&-2[C_Q]^2(R_4)^{ij}_k-2[C_Q]^2(R_4)^{ji}_k-4(R_4)^{ji}_kC_QC_{d_c}
\nonumber\\&&-4(R_4)^{ij}_kC_QC_{d_c}+2(R_4)^{jl}_m(Y_d)_{lk}(Y_d)^{im}C_Q
\nonumber\\&&+2(R_4)^{jl}_m(\Lambda_U)_{knl}(\Lambda_D)^{mni}C_Q-2(R_1)^{jl}_m(\Lambda_U)_{lnk}(\Lambda_D)^{nmi}C_Q
\nonumber\\&&+2(R_9)^{jl}(\Lambda_U)_{lnk}(Y_d)^{in}C_Q-2(R_7)^{jl}(\Lambda_U)_{nlk}(Y_d)^{in}C_Q
\nonumber\\&&+2(R_4)^{jl}_m(Y_d)_{lk}(Y_d)^{im}C_{d_c}+2(R_4)^{jl}_m(\Lambda_D)_{knl}(\Lambda_d)^{mni}C_{d_c}
\nonumber\\&&+2(R_1)^{jl}_m(\Lambda_U)_{lkn}(\Lambda_D)^{nmi}C_{d_c}-2(R_9)^{jl}(\Lambda_U)_{lkn}(Y_d)^{in}C_{d_c}
\nonumber\\&&+2(R_7)^{jl}(\Lambda_U)_{nkl}(Y_d)^{in}C_{d_c}-3(Y_u)^{jl}(\Lambda_U)_{lnk}(\Lambda_D)^{npi}(Y_u)_{qr}(R_1)^{qr}_p
\nonumber\\&&-3(Y_u)^{jl}(\Lambda_U)_{lnk}(Y_d)^{in}(Y_u)_{qr}(R_9)^{qr}-(Y_d)^{jl}(\Lambda_U)_{nlk}(Y_d)^{in}(Y_e)_{qr}(R_5)^{qr}
\nonumber\\&&+(\Lambda_D)^{lmj}(\Lambda_U)_{nlk}(Y_u)^{in}(\Lambda_E)_{rqm}(R_5)^{qr}-(\Lambda_D)^{lmj}(\Lambda_U)_{nlk}(Y_u)^{in}(Y_e)_{mr}(R_2)^{r}
\nonumber\\&&-3(\Lambda_D)^{lmj}(\Lambda_U)_{nlk}(Y_u)^{in}(\Lambda_D)_{rmq}(R_7)^{qr}-3(Y_d)^{jl}(\Lambda_U)_{nlk}(Y_u)^{in}(Y_d)_{qr}(R_7)^{qr}
\nonumber\\&&+2(Y_u)^{jp}(\Lambda_U)_{pnm}(\Lambda_D)^{mli}(\Lambda_D)_{klr}(R_7)^{rn}+2(Y_u)^{jp}(\Lambda_U)_{pnm}(Y_d)^{im}(Y_d)_{rk}(R_7)^{rn}
\nonumber\\&&-2(\Lambda_D)^{pqj}(\Lambda_U)_{npm}(\Lambda_D)^{mli}(\Lambda_D)_{klr}(R_1)^{rn}_q+2(\Lambda_D)^{pqj}(\Lambda_U)_{npm}(Y_d)^{im}(Y_d)_{rk}(R_1)^{rn}_q
\nonumber\\&&-2(Y_d)^{jp}(\Lambda_U)_{nmp}(\Lambda_D)^{mli}(\Lambda_D)_{klr}(R_9)^{rn}-2(Y_d)^{jp}(\Lambda_U)_{npm}(Y_d)^{im}(Y_d)_{rk}(R_9)^{rn}
\nonumber\\&&-2(\Lambda_D)^{qpj}(\Lambda_D)_{mpn}(\Lambda_D)^{mli}(\Lambda_D)_{klr}(R_4)^{nr}_q-2(\Lambda_D)^{qpj}(\Lambda_D)_{mpn}(Y_d)^{im}(Y_d)_{rk}(R_4)^{nr}_q
\nonumber\\&&-2(Y_d)^{jp}(Y_d)_{nm}(\Lambda_D)^{mli}(\Lambda_D)_{klr}(R_4)^{nr}_q-2(Y_d)^{jp}(Y_d)_{nm}(Y_d)^{im}(Y_d)_{rk}(R_4)^{nr}_q
\nonumber\\&&-2(Y_d)^{jp}(Y_e)_{mn}(\Lambda_D)^{lmi}(\Lambda_U)_{rlk}(R_3)^{rn}_q+2(\Lambda_D)^{qpj}(Y_e)_{pn}(Y_d)^{il}(Y_d)_{il}(\Lambda_U)_{rlk}(R_3)^{rn}_q
\nonumber\\&&-2(\Lambda_D)^{qpj}(\Lambda_E)_{npm}(\Lambda_D)^{lmi}(\Lambda_U)_{rlk}(R_3)^{rn}_q+2(\Lambda_D)^{mlj}(\Lambda_D)_{kln}(R_4)^{ni}_mC_Q
\nonumber\\&&+2(Y_d)^{jm}(Y_d)_{nk}(R_4)^{ni}_mC_Q-2(Y_u)^{jl}(\Lambda_U)_{lnk}(R_7)^{in}C_Q
\nonumber\\&&-2(Y_d)^{jr}(\Lambda_U)_{nlk}(R_9)^{in}C_Q-2(\Lambda_D)^{lmj}(\Lambda_U)_{nlk}(R_1)^{in}_mC_Q
\nonumber\\&&-2(R_4)^{lj}_nC_Q(\Lambda_D)^{nqi}(\Lambda_D)_{kql}-2(R_4)^{lj}_nC_Q(Y_d)^{in}(Y_d)_{lk}
\nonumber\\&&-2(R_7)^{jl}C_Q(Y_u)^{iq}(\Lambda_U)_{qlk}-2(R_9)^{jl}C_Q(Y_u)^{iq}(\Lambda_U)_{lqk}
\nonumber\\&&+2(R_1)^{jl}_nC_Q(\Lambda_D)^{qni}(\Lambda_D)_{lqk}+4(R_4)^{ji}_k[C_Q]^2
\nonumber\\&&-(R_4)^{jl}_k(Y_u)_{ln}(Y_e)^{ip}(\gamma_U)^n_p-(R_4)^{jl}_k(Y_u)_{lm}(Y_e)^{im}(\gamma_{H_2})
\nonumber\\&&-(R_4)^{jl}_k(\Lambda_D)_{nml}(\Lambda_D)^{pmi}(\gamma_D)^n_p-(R_4)^{jl}_k(\Lambda_D)_{mnl}(\Lambda_D)^{mpi}(\gamma_L)^n_p
\nonumber\\&&-(R_4)^{jl}_k(Y_d)_{ln}(Y_d)^{ip}(\gamma_D)^n_p+(R_4)^{jl}_k(Y_d)_{lm}(Y_d)^{im}(\gamma_{H_1})
\nonumber\\&&+(R_4)^{jl}_k(\Lambda_D)_{mnl}(Y_d)^{im}(\gamma_{LH_1})^n+(R_4)^{jl}_k(Y_d)_{lm}(\Lambda_D)^{mpi}(\gamma_{LH_1})_p
\nonumber\\&&-2(R_4)^{jl}_kC_Q(\gamma_Q)^i_l-4(R_4)^{ji}_lC_{d_c}(\gamma_D)^l_k+2(\Lambda_D)^{mlj}(\Lambda_D)_{knp}(\gamma_L)^n_l(R_4)^{ip}_m
\nonumber\\&&+2(Y_d)^{jm}(Y_d)_{pk}(\gamma_{H_1})(R_4)^{ip}_m-(\Lambda_D)^{mlj}(Y_d)_{pk}(\gamma_{H_1})(R_4)^{ip}_m
\nonumber\\&&-2(Y_d)^{jm}(\Lambda_D)_{knp}(\gamma_{LH_1})^n(R_4)^{ip}_m-2(Y_u)^{jl}(\Lambda_U)_{npk}(\gamma_U)^n_l(R_7)^{ip}
\nonumber\\&&-2(\Lambda_D)^{lmj}(\Lambda_U)_{pnk}(\gamma_D)^n_l(R_1)^{ip}_m+2(Y_d)^{jl}(\Lambda_U)_{pnk}(\gamma_D)^n_l(R_9)^{ip}
\nonumber\\&&-2(\Lambda_D)^{mlj}(\Lambda_D)_{kln}(R_4)^{in}_p(\gamma_D)^p_m-2(Y_d)^{jm}(Y_d)^{nk}(R_4)^{in}_p(\gamma_D)^p_m
\nonumber\\&&-2(Y_u)^{jl}(\Lambda_U)_{lkn}(R_7)^{in}(\gamma_{H_2})-2(\Lambda_D)^{lmj}(\Lambda_U)_{lnk}(R_1)^{in}_p(\gamma_L)^p_m
\nonumber\\&&-2(\Lambda_D)^{lmj}(\Lambda_U)_{lkn}(R_9)^{in}(\gamma_{LH_1})_m+2(Y_d)^{jl}(\Lambda_U)_{nlk}(R_1)^{in}_m(\gamma_{LH_1})_m
\nonumber\\&&+2(Y_d)^{jl}(\Lambda_U)_{nlk}(R_9)^{in}(\gamma_{H_1})-2(\Lambda_D)^{mlj}(\Lambda_D)_{kln}(\gamma_Q)^n_p(R_4)^{ip}_m
\nonumber\\&&+2(Y_d)^{jm}(Y_d)_{kn}(\gamma_Q)^n_p(R_4)^{ip}_m-2(Y_u)^{jl}(\Lambda_U)_{lkn}(\gamma_D)^n_p(R_7)^{ip}
\nonumber\\&&-2(\Lambda_D)^{lmj}(\Lambda_U)_{nlk}(\gamma_U)^n_p(R_9)^{ip}+2(Y_d)^{jl}(\Lambda_U)_{nlk}(\gamma_U)^n_p(R_9)^{ip}
\nonumber\\&&-2(R_4)^{ji}_l(\Lambda_U)^{mnl}(\Lambda_U)_{mpk}(\gamma_D)^p_n-2(R_4)^{ji}_l(\Lambda_U)^{nml}(\Lambda_U)_{pmk}(\gamma_U)^p_n
\nonumber\\&&-2(R_4)^{ji}_l(\Lambda_D)^{lnm}(\Lambda_D)_{kpm}(\gamma_L)^p_n-2(R_4)^{ji}_l(Y_d)^{ml}(Y_d)_{mk}(\gamma_{H_1})
\nonumber\\&&+2(R_4)^{ji}_l(\Lambda_D)^{lnm}(Y_d)_{mk}(\gamma_{LH_1})_n-2(R_4)^{ji}_l(Y_d)^{nl}(Y_d)_{pk}(\gamma_Q)^p_n
\nonumber\\&&-2(R_4)^{ji}_l(\Lambda_D)^{lmn}(\Lambda_D)_{kmp}(\gamma_Q)^p_n+2(R_4)^{ji}_l(Y_d)^{ml}(\Lambda_D)_{kmp}(\gamma_{LH_1})^p
\nonumber\\&&-2(R_4)^{ji}_k[\frac{33}{75}g^4_1-4g^4_3]+4(R_4)^{ji}_k[\frac{33}{300}g^4_1+\frac{3}{4}g^4_2+12g^4_3]+j\longleftrightarrow k
\end{eqnarray}

\begin{eqnarray}
(16\pi^2)^2(\beta^{(2)}_{R_5})^{ij}&=&
+6(\Lambda_D)^{lip}(Y_u)_{pn}C_{u^c}(R_3)^{jn}_l+6(\Lambda_D)^{nip}(Y_u)_{pm}(R_3)^{jm}_nC_{d_c}
\nonumber\\&&-12(\Lambda_D)^{mil}(Y_u)_{ln}(\Lambda_D)_{mpq}(Y_u)^{pn}(R_5)^{jp}+12(\Lambda_D)^{mil}(Y_u)_{ln}(Y_d)_{pm}(Y_u)^{pm}(R_2)^{j}
\nonumber\\&&+12(\Lambda_D)^{mil}(Y_u)_{ln}(\Lambda_U)_{qpm}(\Lambda_U)^{nrp}(R_3)^{qj}_r-4(R_5)^{ij}[C_{H_2}]^2
\nonumber\\&&-2[C_L]^2(R_5)^{ij}-2[C_{e^c}]^2(R_5)^{ij}-4(R_5)^{ij}C_{e^c}C_{H_2}
\nonumber\\&&+6(R_3)^{jl}_m(Y_u)_{nl}(\Lambda_D)_{min}C_L+6(R_3)^{jl}_m(Y_u)_{nl}(\Lambda_D)^{min}C_{H_2}
\nonumber\\&&+12(Y_e)^{ij}(Y_u)_{mn}(\Lambda_U)^{nrq}(Y_u)_{pr}(R_4)^{mp}_q+6(Y_e)^{ij}(Y_u)_{mn}(\Lambda_D)^{nqr}(Y_u)_{rp}(R_1)^{mp}_q
\nonumber\\&&-6(Y_e)^{ij}(Y_u)_{mn}(Y_d)_{rn}(Y_u)_{rp}(R_9)^{mp}-12(\Lambda_E)^{jqp}(\Lambda_D)_{mpn}(\Lambda_D)^{mil}(Y_u)_{lr}(R_1)^{nr}_q
\nonumber\\&&+12(Y_e)^{jp}(\Lambda_D)_{mnp}(\Lambda_D)^{mil}(Y_u)_{lr}(R_9)^{nr}+12(Y_e)^{qj}(Y_d)_{nm}(\Lambda_D)^{mil}(Y_u)_{lr}(R_1)^{nr}_q
\nonumber\\&&+6(\Lambda_D)^{mil}(Y_u)_{ln}(R_3)^{nj}_mC_{e^c}-6(R_3)^{lj}_nC_{e_c}(\Lambda_D)^{niq}(Y_u)_{ql}
\nonumber\\&&+4(R_5)^{ji}[C_L]^2-2(R_5)^{il}(\Lambda_E)_{lnm}(\Lambda_E)^{jpm}(\gamma_L)^n_p
\nonumber\\&&-2(R_5)^{il}(\Lambda_E)_{lnm}(Y_e)^{mj}(\gamma_{LH_1})^n-2(R_5)^{il}(Y_e)_{nl}(Y_e)^{pj}(\gamma_L)^n_p
\nonumber\\&&-(R_5)^{lj}(\Lambda_E)_{nml}(\Lambda_E)^{pmi}(\gamma_E)^n_p-(R_5)^{lj}(Y_e)_{ln}(Y_e)^{ip}(\gamma_E)^n_p
\nonumber\\&&-(R_5)^{lj}(\Lambda_E)_{mnl}(\Lambda_E)^{mpi}(\gamma_L)^n_p-(R_5)^{lj}(Y_e)_{lm}(Y_e)^{im}(\gamma_{H_1})
\nonumber\\&&-2(R_5)^{il}(Y_e)_{ml}(Y_e)^{mj}(\gamma_{H_1})-3(R_5)^{lj}(\Lambda_D)_{mln}(\Lambda_D)^{mip}(\gamma_Q)^n_p
\nonumber\\&&-3(R_5)^{lj}(\Lambda_D)_{nlm}(\Lambda_D)^{pim}(\gamma_D)^n_p-2(R_5)^{ij}C_{e^c}(\gamma_E)^j_l
\nonumber\\&&-2(R_5)^{ij}C_L(\gamma_L)^i_l-(R_5)^{ij}[C_{H_2}\gamma_{H_2}]
\nonumber\\&&-6(\Lambda_D)^{lim}(Y_u)_{np}(\gamma_Q)^n_l(R_3)^{pj}_m+6(\Lambda_D)^{mil}(Y_u)_{ln}(R_3)^{jn}_p(\gamma_D)^p_m
\nonumber\\&&+6(\Lambda_D)^{mil}(Y_u)_{ln}(\gamma_U)^n_p(R_3)^{pj}_m-3(R_5)^{ij}(Y_u)_{mn}(Y_u)^{mp}(\gamma_U)^p_n
\nonumber\\&&-3(R_5)^{ij}(Y_u)_{nm}(Y_u)^{pm}(\gamma_Q)^p_n+4(R_5)^{im}[C_{e^c}]^j_m
\nonumber\\&&-2(R_5)^{ij}[\frac{99}{100}g^4_1+\frac{3}{4}g^4_2]+2(R_5)^{ij}[\frac{99}{25}g^4_1]
\nonumber\\&&+2(R_5)^{ij}[\frac{99}{100}g^4_1+\frac{3}{4}g^4_2]
\end{eqnarray}

\begin{eqnarray}
(16\pi^2)^2(\beta^{(2)}_{R_7})^{ij}&=&
+2(Y_u)^{ip}(Y_u)_{np}C_Q(R_5)^{nj}+4(\Lambda_U)^{pjl}(Y_u)_{np}C_Q(R_4)^{in}_l
\nonumber\\&&+2(\Lambda_D)^{jlp}(Y_u)_{pn}C_{u^c}(R_1)^{in}_l-2(Y_d)^{pi}(Y_u)_{pn}C_{u^c}(R_9)^{in}
\nonumber\\&&+2(Y_u)^{ip}(Y_u)_{mp}(R_7)^{mj}C_{H_2}+4(\Lambda_D)^{pjn}(Y_u)_{mp}(R_4)^{im}_nC_{d_c}
\nonumber\\&&+2(\Lambda_D)^{jnp}(Y_u)_{pm}(R_1)^{im}_nC_L-2(Y_d)^{pj}(Y_u)_{pm}(R_9)^{im}C_{H_1}
\nonumber\\&&-2(Y_u)^{il}(Y_u)_{nl}(Y_u)_{qp}(Y_u)^{np}(R_7)^{qj}+4(\Lambda_U)^{lmj}(Y_u)_{nl}(\Lambda_U)_{qpm}(\Lambda_D)^{prn}(R_1)^{iq}_r
\nonumber\\&&-4(\Lambda_U)^{lmj}(Y_u)_{nl}(\Lambda_U)_{qpm}(Y_d)^{np}(R_9)^{iq}-4(\Lambda_D)^{jml}(Y_u)_{ln}(\Lambda_D)_{pmq}(\Lambda_U)^{nrp}(R_4)^{iq}_r
\nonumber\\&&-2(Y_u)^{jl}(Y_u)_{ln}(Y_d)_{qp}(\Lambda_U)^{npr}(R_4)^{iq}_r-4(R_7)^{ij}[C_{H_2}]^24
\nonumber\\&&-2[C_Q]^2(R_7)^{ji}-2[C_{d_c}]^2(R_7)^{ij}-4(R_7)^{ij}C_{d_c}C_{H_2}
\nonumber\\&&-4(R_7)^{ji}C_QC_{H_2}-4(R_4)^{il}_m(Y_u)_{ln}(\Lambda_U)^{nmj}C_{d_c}
\nonumber\\&&-2(R_9)^{il}(Y_u)_{nl}(Y_d)^{nj}C_{d_c}-2(R_1)^{il}_m(Y_u)_{nl}(\Lambda_D)^{jmn}C_{d_c}
\nonumber\\&&+2(R_7)^{lj}(Y_u)_{ln}(Y_u)^{in}C_Q+2(R_4)^{il}_m(Y_u)_{ln}(\Lambda_U)^{nmj}C_{H_2}
\nonumber\\&&-2(R_9)^{il}(Y_u)_{nl}(Y_d)^{nj}C_{H_2}+2(R_1)^{il}_m(Y_u)_{nl}(\Lambda_D)^{jmn}C_{H_2}
\nonumber\\&&-2(R_7)^{lj}(Y_u)_{ln}(Y_u)^{in}C_{H_2}-3(Y_u)^{il}(Y_u)_{nl}(Y_u)^{jn}(Y_u)_{qr}(R_9)^{qr}
\nonumber\\&&+3(Y_u)^{il}(Y_u)_{nl}(\Lambda_D)^{jpn}(Y_u)_{qr}(R_1)^{qr}_p+3(\Lambda_D)^{jml}(Y_u)_{ln}(Y_u)^{in}(\Lambda_d)^{rmq}(R_7)^{qr}
\nonumber\\&&+3(Y_d)^{lj}(Y_u)_{ln}(Y_u)^{in}(Y_u)_{qr}(R_7)^{qr}-(\Lambda_D)^{jml}(Y_u)_{ln}(Y_u)^{in}(\Lambda_E)^{rqm}(R_5)^{qr}
\nonumber\\&&-(Y_d)^{lj}(Y_u)_{ln}(Y_u)^{in}(Y_e)_{qr}(R_5)^{qr}-(\Lambda_D)^{jml}(Y_u)_{ln}(Y_u)^{in}(Y_e)_{mr}(R_2)^{r}
\nonumber\\&&+4(\Lambda_D)^{pqi}(\Lambda_E)_{npm}(\Lambda_U)^{lmj}(Y_u)_{rl}(R_1)^{rn}_q-4(Y_d)^{ip}(\Lambda_U)_{npm}(\Lambda_U)^{lmj}(Y_u)_{rl}(R_1)^{rn}_q
\nonumber\\&&-2(\Lambda_D)^{qpi}(\Lambda_E)_{npm}(\Lambda_D)^{jml}(Y_u)_{lr}(R_3)^{rn}_q-2(\Lambda_D)^{qpi}(Y_e)_{pn}(Y_d)^{lj}(Y_u)_{lr}(R_3)^{rn}_q
\nonumber\\&&-2(Y_d)^{iq}(Y_e)_{mn}(\Lambda_D)^{jml}(Y_u)_{lr}(R_3)^{rn}_q+4(\Lambda_D)^{qpi}(\Lambda_D)_{mpn}(\Lambda_U)^{lmj}(Y_e)_{il}(R_4)^{nr}_q
\nonumber\\&&+4(Y_d)^{iq}(Y_d)_{nm}(\Lambda_U)^{lmj}(Y_u)_{rl}(R_4)^{nr}_q-4(\Lambda_D)^{jqp}(Y_u)_{pn}(Y_u)^{il}(Y_u)_{rl}(R_1)^{rn}_q
\nonumber\\&&+2(Y_d)^{pj}(Y_u)_{pn}(Y_u)^{il}(Y_u)_{rl}(R_9)^{rn}+2(Y_u)^{il}(Y_u)_{nl}(R_7)^{nj}C_{d_c}
\nonumber\\&&+4(\Lambda_U)^{lmj}(Y_u)_{nl}(R_4)^{ni}_mC_Q+2(\Lambda_D)^{jml}(Y_u)_{ln}(R_1)^{in}_mC_Q
\nonumber\\&&-2(Y_d)^{lj}(Y_u)_{ln}(R_9)^{in}C_Q-2(R_1)^{il}_nC_Q(\Lambda_D)^{jnq}(Y_u)_{ql}
\nonumber\\&&+2(R_9)^{il}C_Q(Y_d)^{qj}(Y_u)_{ql}+4(R_4)^{li}_nC_Q(\Lambda_U)^{qnj}(Y_u)_{lq}
\nonumber\\&&+2(R_7)^{lj}C_{d_c}(Y_u)^{iq}(Y_u)_{lq}+4(R_7)^{ij}[C_{d_c}]^2+4(Y_u)^{ij}[C_Q]^2
\nonumber\\&&-2(R_7)^{il}(Y_d)_{ml}(Y_d)^{jm}(\gamma_{H_1})+2(R_7)^{il}(Y_d)_{ml}(\Lambda_D)^{jpm}(\gamma_{LH_1})_p
\nonumber\\&&-2(R_7)^{il}(\Lambda_D)_{lnm}(\Lambda_D)^{jpm}(\gamma_L)^n_p-2(R_7)^{il}(\Lambda_U)_{mnl}(\Lambda_U)^{mpj}(\gamma_D)^n_p
\nonumber\\&&-2(R_7)^{il}(\Lambda_U)_{nml}(\Lambda_U)^{pmj}(\gamma_U)^n_p-(R_7)^{lj}(Y_u)_{lm}(Y_u)^{im}(\gamma_{H_2})
\nonumber\\&&-(R_7)^{lj}(Y_u)_{ln}(Y_u)^{ip}(\gamma_U)^n_p-2(Y_u)^{il}C_{d_c}(\gamma_D)^j_l
\nonumber\\&&-2(Y_u)^{lj}C_Q(\gamma_Q)^i_l-2(R_7)^{ij}[C_{H_2}\gamma_{H_2}]
\nonumber\\&&+2(Y_u)^{il}(Y_u)_{pn}(\gamma_U)^n_l(R_7)^{pj}+2(\Lambda_D)^{jml}(Y_u)_{np}(\gamma_Q)^n_l(R_1)^{ip}_n
\nonumber\\&&-2(Y_d)^{lj}(Y_u)_{np}(\gamma_Q)^n_l(R_9)^{ip}+4(\Lambda_U)^{lmj}(Y_u)_{pn}(\gamma_U)^n_l(R_4)^{ip}_m
\nonumber\\&&+2(Y_u)^{il}(Y_u)_{nl}(R_7)^{nj}(\gamma_{H_2})+4(\Lambda_U)^{lmj}(Y_u)_{nl}(R_4)^{ni}_p(\gamma_D)^p_m
\nonumber\\&&+2(\Lambda_D)^{jml}(Y_u)_{ln}(R_1)^{in}_p(\gamma_L)^p_m+2(\Lambda_D)^{jml}(Y_u)_{ln}(R_9)^{in}(\gamma_{LH_1})_m
\nonumber\\&&-2(Y_d)^{nj}(Y_u)_{ln}(R_1)^{in}_p(\gamma_{LH_1})^p-2(Y_d)^{jl}(Y_u)_{ln}(R_9)^{in}(\gamma_{H_1})
\nonumber\\&&+2(Y_u)^{il}(Y_u)_{nl}(\gamma_Q)^n_p(R_7)^{pj}+2(\Lambda_d)^{jml}(Y_u)_{ln}(\gamma_U)^n_p(R_1)^{ip}_m
\nonumber\\&&-2(Y_d)^{jl}(Y_u)_{ln}(\gamma_U)^n_p(R_9)^{ip}+4(\Lambda_U)^{lmj}(Y_u)_{nl}(\gamma_Q)^n_p(R_4)^{ip}_m
\nonumber\\&&-3(R_7)^{ij}(Y_u)^{mn}(Y_u)_{mp}(\gamma_U)^p_n-3(R_7)^{ij}(Y_u)^{nm}(Y_u)_{pm}(\gamma_Q)^p_n
\nonumber\\&&-2(R_7)^{ij}[\frac{33}{100}g^4_1+\frac{3}{4}g^4_2]+2(R_7)^{ij}[\frac{99}{25}g^4_1]
\nonumber\\&&+2(R_7)^{ij}[\frac{33}{100}g^4_1+\frac{3}{4}g^4_2]
\end{eqnarray}

\begin{eqnarray}
(16\pi^2)^2(\beta^{(2)}_{R_9})^{ij}&=&
+2(Y_d)^{ip}(Y_u)_{np}C_Q(R_9)^{nj}-2(\Lambda_D)^{pil}(Y_d)_{np}C_Q(R_1)^{nj}_l
\nonumber\\&&-2(\Lambda_D)^{lpi}(Y_e)_{pn}C_{e^c}(R_3)^{jn}_l-2(Y_u)^{pj}(Y_d)_{pn}C_{d_c}(R_7)^{in}
\nonumber\\&&-4(\Lambda_U)^{jlp}(Y_e)_{np}C_Q(R_4)^{in}_l+2(Y_d)^{ip}(Y_d)_{mp}(R_9)^{mj}C_{H_1}
\nonumber\\&&-2(\Lambda_D)^{lpi}(Y_e)_{pm}(R_3)^{jm}_lC_{d_c}-2(Y_u)^{pj}(Y_d)_{pm}(R_7)^{im}C_{H_2}
\nonumber\\&&-4(\Lambda_U)^{jlp}(Y_d)_{mp}(R_4)^{im}_lC_{d_c}+2(\Lambda_D)^{lmi}(Y_d)_{nl}(\Lambda_D)_{pmq}(\Lambda_d)^{prn}(R_1)^{qj}_r
\nonumber\\&&+2(Y_d)^{il}(Y_d)_{nl}(Y_d)_{qp}(\Lambda_D)^{prn}(R_1)^{qj}_r-2(\Lambda_D)^{lmi}(Y_d)_{nl}(\Lambda_D)_{pmq}(Y_d)^{np}(R_9)^{qj}
\nonumber\\&&+2(Y_d)^{il}(Y_d)_{nl}(Y_d)_{qp}(Y_d)^{np}(R_9)^{qj}-2(\Lambda_D)^{lmi}(Y_d)_{nl}(\Lambda_E)_{qpm}(\Lambda_D)^{rpn}(R_3)^{jq}_r
\nonumber\\&&-2(\Lambda_D)^{lmi}(Y_d)_{nl}(Y_e)_{mq}(Y_d)^{nr}(R_3)^{jq}_r+2(Y_e)^{in}(Y_d)_{nl}(Y_e)_{pq}(\Lambda_D)^{rpn}(R_3)^{jq}_r
\nonumber\\&&+4(\Lambda_U)^{jml}(Y_d)_{ln}(\Lambda_D)_{mpq}(\Lambda_D)^{rpn}(R_4)^{iq}_r+4(\Lambda_U)^{jml}(Y_d)_{nl}(Y_d)_{qm}(Y_d)^{nr}(R_4)^{iq}_r
\nonumber\\&&+4(\Lambda_U)^{jml}(Y_d)_{nl}(\Lambda_U)^{pqm}(Y_u)^{np}(R_7)^{iq}+2(Y_u)^{lj}(Y_d)_{ln}(Y_u)_{pq}(\Lambda_d)^{nrp}(R_1)^{iq}_r
\nonumber\\&&-4(\Lambda_U)^{jml}(Y_d)_{nl}(\Lambda_U)^{qpm}(\Lambda_D)^{prn}(R_1)^{iq}_r+4(\Lambda_U)^{jml}(Y_d)_{nl}(\Lambda_U)^{qpm}(Y_d)^{np}(R_9)^{iq}
\nonumber\\&&-2(Y_u)^{lj}(Y_d)_{ln}(Y_u)_{pq}(Y_d)^{pn}(R_9)^{iq}-4(R_9)^{ij}[C_{H_1}]^2
\nonumber\\&&-2[C_Q]^2(R_9)^{ij}-2[C_{u^c}]^2(R_9)^{ij}-4(R_9)^{ij}C_{u^c}C_{H_1}
\nonumber\\&&-(R_9)^{ij}C_QC_{H_1}-2(R_7)^{il}(Y_d)_{nl}(Y_u)^{nj}C_{u^c}
\nonumber\\&&+4(R_4)^{il}_m(Y_d)_{ln}(\Lambda_U)^{jmn}C_{u^c}-(R_9)^{lj}(Y_d)_{ln}(Y_d)^{in}C_Q
\nonumber\\&&+2(R_1)^{lj}_m(Y_d)_{ln}(\Lambda_D)^{nmi}C_Q-2(R_3)^{jl}_m(Y_e)_{nl}(\Lambda_D)^{mni}C_Q
\nonumber\\&&+4(R_4)^{il}_m(Y_d)_{ln}(\Lambda_U)^{jmn}C_{H_1}-2(R_7)^{il}(Y_d)_{nl}(Y_u)^{nj}C_{H_1}
\nonumber\\&&+2(R_9)^{lj}(Y_d)_{ln}(Y_d)^{in}C_{H_1}-2(R_1)^{lj}_m(Y_d)_{ln}(\Lambda_D)^{nmi}C_{H_1}
\nonumber\\&&-2(R_3)^{jl}_m(Y_e)_{nl}(\Lambda_D)^{mni}C_{H_1}+12(Y_u)^{ij}(Y_u)_{mn}(\Lambda_U)^{nrq}(Y_d)_{pr}(R_4)^{mp}_q
\nonumber\\&&-6(Y_u)^{ij}(Y_u)_{mn}(Y_u)^{rn}(Y_d)_{rp}(R_7)^{mp}-6(Y_u)^{ij}(Y_u)_{nm}(\Lambda_D)^{rqn}(Y_d)_{pr}(R_1)^{pm}_q
\nonumber\\&&+6(Y_u)^{ij}(Y_u)_{nm}(Y_d)^{nr}(Y_d)_{pr}(R_9)^{pm}-6(Y_u)^{ij}(Y_u)_{nm}(\Lambda_D)^{qrn}(Y_e)_{rp}(R_3)^{np}_q
\nonumber\\&&-(\Lambda_D)^{lmi}(Y_d)_{nl}(Y_u)^{nj}(\Lambda_E)_{qmr}(R_5)^{rq}-(Y_d)^{il}(Y_d)_{nl}(Y_u)^{nj}(Y_e)_{rq}(R_5)^{rq}
\nonumber\\&&-(\Lambda_D)^{lmi}(Y_u)_{nl}(Y_u)^{nj}(Y_e)_{rq}(R_2)^{q}-(Y_d)^{il}(Y_d)_{nl}(Y_u)^{nj}(Y_d)_{qr}(R_7)^{qr}
\nonumber\\&&-(\Lambda_D)^{lmi}(Y_d)_{nl}(Y_u)^{nj}(\Lambda_D)_{rmq}(R_7)^{qr}-3(Y_u)^{lj}(Y_d)_{ln}(\Lambda_D)^{npi}(Y_u)_{qr}(R_1)^{qr}_p
\nonumber\\&&+3(Y_u)^{lj}(Y_d)_{ln}(Y_d)_{in}(Y_u)^{qr}(R_7)^{qr}-4(\Lambda_D)^{pqi}(\Lambda_D)_{npm}(\Lambda_U)^{jml}(Y_d)_{rl}(R_1)^{rn}_q
\nonumber\\&&+4(Y_d)^{ip}(\Lambda_U)_{nmp}(\Lambda_U)^{jlm}(Y_d)_{rl}(R_9)^{rn}+4(Y_d)^{ip}(\Lambda_U)_{nmp}(\Lambda_U)^{jlm}(Y_d)_{rl}(R_9)^{rn}
\nonumber\\&&+4(Y_u)^{ip}(\Lambda_U)_{npm}(\Lambda_U)^{jml}(Y_d)_{rl}(R_5)^{rn}+2(Y_u)^{pj}(\Lambda_D)_{nmp}(\Lambda_d)^{lmi}(Y_d)_{rl}(R_5)^{rn}
\nonumber\\&&+2(Y_u)^{pj}(Y_d)_{pn}(Y_d)^{il}(Y_d)_{rl}(R_5)^{rn}+4(Y_u)^{pj}(\Lambda_D)_{mnp}(\Lambda_D)^{mli}(Y_e)_{lr}(R_5)^{nr}
\nonumber\\&&-4(Y_u)^{pj}(Y_d)_{pm}(\Lambda_D)^{mli}(Y_e)_{lr}(R_2)^{r}-18(Y_u)^{ij}(Y_u)_{ml}(R_9)^{ml}C_{H_1}
\nonumber\\&&+2(\Lambda_D)^{lmi}(Y_d)_{nl}(R_1)^{nj}_mC_{u^c}+2(Y_d)^{il}(Y_d)_{nl}(R_9)^{jp}C_{u^c}
\nonumber\\&&-2(\Lambda_D)^{mli}(Y_e)_{ln}(R_3)^{jn}_mC_{u^c}-2(Y_u)^{lj}(Y_d)_{ln}(R_7)^{in}C_Q
\nonumber\\&&-4(\Lambda_U)^{jlm}(Y_d)_{nl}(R_4)^{ni}_mC_Q-4(R_4)^{li}_nC_Q(\Lambda_U)^{jqn}(Y_d)_{lq}
\nonumber\\&&+2(R_7)^{in}C_Q(Y_u)^{qj}(Y_d)_{ql}+2(R_1)^{lj}_nC_{u^c}(\Lambda_D)^{qni}(Y_d)_{lq}
\nonumber\\&&+2(R_9)^{lj}C_{u^c}(Y_d)^{iq}(Y_d)_{lq}+2(R_3)^{jl}_nC_{u^c}(\Lambda_D)^{iqn}(Y_e)_{ql}
\nonumber\\&&+4(R_9)^{ij}[C_{u^c}]^2+4(R_9)^{ij}[C_Q]^2-2(R_9)^{il}(\Lambda_U)_{lmn}(\Lambda_U)^{jmp}(\gamma_D)^n_p
\nonumber\\&&-2(R_9)^{il}(Y_u)_{nl}(Y_u)^{pj}(\gamma_Q)^n_p-2(R_9)^{il}(Y_u)_{ml}(Y_u)^{mj}(\gamma_{H_2})
\nonumber\\&&-(R_9)^{lj}(Y_d)_{lm}(Y_d)^{im}(\gamma_{H_1})-(R_9)^{lj}(Y_d)_{ln}(Y_d)^{ip}(\gamma_D)^n_p
\nonumber\\&&+(R_9)^{lj}(\Lambda_D)_{nml}(\Lambda_D)^{imp}(\gamma_D)^n_p-(R_9)^{lj}(\Lambda_D)_{mnl}(\Lambda_D)^{mpi}(\gamma_L)^n_p
\nonumber\\&&-(R_9)^{lj}(Y_u)_{ln}(Y_u)^{ip}(\gamma_U)^n_p-(R_9)^{lj}(Y_u)_{lm}(Y_u)^{im}(\gamma_{H_2})
\nonumber\\&&+(R_9)^{lj}(\Lambda_D)_{mnl}(Y_d)^{im}(\gamma_{LH_1})^n-6(Y_u)_{ml}(Y_u)^{ij}(R_7)^{mp}(\gamma_U)^l_p
\nonumber\\&&-6(Y_u)_{lm}(Y_u)^{ij}(R_7)^{pm}(\gamma_Q)^l_p-2(R_9)^{il}C_{u^c}(\gamma_U)^j_l
\nonumber\\&&+(R_9)^{lj}(Y_d)_{lm}(\Lambda_D)^{mpi}(\gamma_{LH_1})_p-2(R_9)^{ij}C_Q(\gamma_Q)^i_l
\nonumber\\&&-2(R_9)^{ij}[C_{H_1}\gamma_{H_1}]+2(Y_d)^{il}(Y_d)_{pn}(\gamma_D)^n_l(R_9)^{pj}
\nonumber\\&&-2(\Lambda_D)^{lmi}(Y_d)_{pn}(\gamma_D)^n_l(R_1)^{pj}_m+2(\Lambda_D)^{mli}(Y_e)_{np}(\gamma_L)^n_l(R_3)^{jp}_m
\nonumber\\&&+2(Y_e)^{im}(Y_e)_{np}(\gamma_{LH_1})^n(R_3)^{jp}_m-2(Y_u)^{lj}(Y_d)_{np}(\gamma_Q)^n_l(R_7)^{ip}
\nonumber\\&&-4(\Lambda_U)^{jml}(Y_d)_{np}(\gamma_D)^n_p(R_4)^{ip}_m+2(Y_d)^{il}(Y_d)_{nl}(R_9)^{nj}(\gamma_{H_1})
\nonumber\\&&-2(\Lambda_D)^{lmi}(Y_d)_{nl}(R_1)^{nj}_p(\gamma_L)^p_m-2(\Lambda_D)^{mli}(Y_e)_{ln}(R_3)^{jp}_p(\gamma_D)^p_m
\nonumber\\&&-2(\Lambda_D)^{lmi}(Y_d)_{nl}(R_9)^{nj}(\gamma_{LH_1})_m+2(Y_d)^{il}(Y_d)_{nl}(R_1)^{nj}_p(\gamma_{LH_1})^p
\nonumber\\&&-2(Y_u)^{lj}(Y_d)_{nl}(R_7)^{in}(\gamma_{H_2})-4(\Lambda_U)^{jlm}(Y_d)_{nl}(R_1)^{in}_p(\gamma_D)^p_m
\nonumber\\&&-2(\Lambda_D)^{lmi}(Y_d)_{nl}(\gamma_Q)^n_p(R_1)^{pj}_m+2(Y_d)^{il}(Y_d)_{nl}(\gamma_Q)^n_p(R_9)^{pj}
\nonumber\\&&-2(\Lambda_D)^{mli}(Y_e)_{ln}(\gamma_E)^n_p(R_3)^{jp}_m-2(Y_u)^{lj}(Y_d)_{ln}(\gamma_D)^n_p(R_7)^{ip}
\nonumber\\&&-4(\Lambda_U)^{jml}(Y_d)_{nl}(\gamma_Q)^n_p(R_4)^{ip}_m-3(R_9)^{ij}(Y_d)^{mn}(Y_d)_{mp}(\gamma_D)^p_n
\nonumber\\&&-3(R_9)^{ij}(Y_d)^{nm}(Y_d)_{pm}(\gamma_Q)^p_n-(R_1)^{ij}_l(\Lambda_E)^{mnl}(Y_e)_{pm}(\gamma_L)^p_n
\nonumber\\&&-(R_1)^{ij}_l(\Lambda_E)^{nml}(Y_e)_{mp}(\gamma_E)^p_n+(R_1)^{ij}_l(Y_e)^{lm}(Y_e)_{pm}(\gamma_{LH_1})^p
\nonumber\\&&-2(R_9)^{ij}[\frac{99}{100}g^4_1+\frac{3}{4}g^4_2]+2(R_9)^{ij}[\frac{132}{25}g^4_1+12g^4_3]
\nonumber\\&&+2(R_9)^{ij}[\frac{33}{300}g^4_1+\frac{3}{4}g^4_2+12g^4_3]
\end{eqnarray}

 Two-loop $\beta$-functions for the various $\phi \phi ^*$ mass terms are given by:

\begin{eqnarray}
(16\pi^2)^2(\beta^{(2)}_{m^2_Q})^i_j&=&(16\pi^2)^2\beta{}_{m^2_Q}^{(2)MSSM}
-2(Y_u)_{jk}(Y_u)^{pq}(R_1)^{ik}_l(R_1)^l_{pq}-2(Y_u)_{jk}(Y_u)^{pq}(R_9)^{ik}(R_9)_{pq}
\nonumber\\&&-2(\Lambda_D)_{kmj}(\Lambda_D)^{qmp}(R_7)^{ik}(R_7)_{pq}-2(Y_d)_{jk}(Y_d)^{pq}(R_7)^{ik}(R_7)_{pq}
\nonumber\\&&+4(\Lambda_D)^{lki}(\Lambda_D)_{mkp}(R_4)^{pn}_l(R_4)^m_{jn}+4(Y_d)^{il}(Y_d)_{pm}(R_4)^{pn}_l(R_4)^m_{jn}
\nonumber\\&&+2(Y_u)^{ik}(Y_u)_{pk}(R_7)^{pn}(R_7)_{jn}-2(\Lambda_D)^{kli}(Y_u)_{pk}(R_1)^{pn}_l(R_9)_{jn}
\nonumber\\&&-2(Y_d)^{ik}(\Lambda_D)_{kmp}(R_9)^{pn}(R_1)^m_{jn}+2(Y_d)^{ik}(Y_d)_{pk}(R_9)^{pn}(R_9)_{jn}
\nonumber\\&&-2C_{H_2}(R_7)^{ik}(R_7)_{jk}+4C_{d_c}(R_4)^{ik}_l(R_1)^l_{kj}
\nonumber\\&&-2C_{H_1}(R_9)^{ik}(R_9)_{jk}+2C_L(R_1)^{ik}_l(R_1)^l_{kj}
\nonumber\\&&-2C_{u^c}(R_1)^{ik}_l(R_1)^l_{jk}-2C_{u^c}(R_9)^{ik}(R_9)_{jk}
\nonumber\\&&-2C_{d_c}(R_7)^{ik}(R_7)_{jk}-4C_Q(R_4)^{ik}_l(R_4)^l_{kj}
\nonumber\\&&-2(R_4)^{ik}_l(R_4)^l_{jm}(\gamma_Q)^m_k-(R_9)^{ik}(R_9)_{jm}(\gamma_U)^m_k
\nonumber\\&&-(R_7)^{ik}(R_7)_{jm}(\gamma_D)^m_k-(R_1)^{ik}_l(R_1)^l_{jm}(\gamma_U)^m_k
\nonumber\\&&-2(R_4)^{ik}_l(R_4)^m_{jk}(\gamma_Q)^l_m-(R_7)^{ik}(R_7)_{jk}(\gamma_{H_2})
\nonumber\\&&-(R_9)^{ik}(R_9)_{jk}(\gamma_{H_1})-(R_1)^{ik}_l(R_1)^m_{jk}(\gamma_L)^l_m
\end{eqnarray}

\begin{eqnarray}
(16\pi^2)^2(\beta^{(2)}_{m^2_{u^c}})^i_j&=&(16\pi^2)^2\beta{}_{m^2_{u^c}}^{(2)MSSM}
-2(Y_u)_{kj}(Y_u)^{qp}(R_9)^{ki}(R_9)_{qp}-2(Y_u)_{kj}(Y_u)^{qp}(R_1)^{ki}_l(R_1)^l_{pq}
\nonumber\\&&+8(\Lambda_U)^{ilk}(\Lambda^D)_{kmp}(R_4)^{pn}_l(R_1)^m_{nj}-8(\Lambda_U)^{ilk}(\Lambda^D)_{kmp}(R_4)^{pm}_l(R_9)_{nj}
\nonumber\\&&-4(Y_u)^{ki}(Y_d)_{kp}(R_7)^{np}(R_9)_{nj}-2(Y_u)^{ki}(\Lambda^D)_{pmk}(R_7)^{np}(R_1)^m_{nj}
\nonumber\\&&-2C_{d_c}(R_3)^{ik}_l(R_3)^l_{jk}-4C_L(R_1)^{ki}_l(R_1)^l_{kj}
\nonumber\\&&-4C_{H_1}(R_9)^{ki}(R_9)_{kj}-2C_{e^c}(R_3)^{ik}_l(R_3)^l_{kj}
\nonumber\\&&-4C_Q(R_1)^{ki}_l(R_1)^l_{kj}-4C_Q(R_9)^{ki}(R_9)_{kj}
\nonumber\\&&-2(R_1)^{ki}_l(R_1)^l_{mj}(\gamma_Q)^m_k-2(R_9)^{ki}(R_9)_{mj}(\gamma_Q)^m_k
\nonumber\\&&-2(R_3)^{ik}_l(R_3)^l_{jm}(\gamma_E)^m_k-2(R_1)^{ki}_l(R_1)^l_{kj}(\gamma_L)^l_m
\nonumber\\&&-2(R_9)^{ki}(R_9)_{kj}(\gamma_{H_1})-2(R_3)^{ik}_l(R_3)^l_{jk}(\gamma_D)^l_m
\end{eqnarray}

\begin{eqnarray}
(16\pi^2)^2(\beta^{(2)}_{m^2_{d^c}})^i_j&=&(16\pi^2)^2\beta{}_{m^2_{d^c}}^{(2)MSSM}
-8(\Lambda_U)^{lik}(Y_u)_{nl}(R_4)^{mn}_j(R_7)_{mk}-4(Y_d)^{li}(Y_u)_{ln}(R_3)^{nm}_j(R_2)_{m}
\nonumber\\&&+4(\Lambda_D)^{ikl}(Y_u)_{ln}(R_3)^{nm}_j(R_5)_{km}-6(\Lambda_D)^{jmk}(\Lambda_D)_{mqp}(R_7)^{ki}(R_5)_{qp}
\nonumber\\&&-6(Y_d)_{kj}(Y_d)^{qp}(R_7)^{ki}(R_7)_{qp}+8(\Lambda_U)^{kil}(Y_u)_{pk}(R_4)^{pn}_l(R_7)_{nj}
\nonumber\\&&+4C_{d_c}(R_3)^{lk}_j(R_1)^i_{lk}+16C_{d_c}(R_4)^{kl}_j(R_4)^i_{kl}
\nonumber\\&&-4C_{H_2}(R_7)^{ki}(R_7)_{kj}-4C_Q(R_7)^{ki}(R_7)_{kj}
\nonumber\\&&-(R_3)^{kl}_j(R_3)^i_{km}(\gamma_E)^m_l-(R_3)^{lk}_j(R_3)^i_{mk}(\gamma_U)^m_l
\nonumber\\&&-4(R_3)^{kl}_j(R_3)^i_{km}(\gamma_Q)^m_l-2(R_7)^{ki}(R_7)_{mj}(\gamma_Q)^m_k
\nonumber\\&&-2(R_7)^{ki}(R_7)_{kj}(\gamma_{H_2})
\end{eqnarray}

\begin{eqnarray}
(16\pi^2)^2(\beta^{(2)}_{m^2_L})^i_j&=&(16\pi^2)^2\beta{}_{m^2_L}^{(2)MSSM}
+12(\Lambda_D)^{kil}(\Lambda_U)_{npl}(R_1)^{mn}_j(R_4)^p_{km}+6(\Lambda_D)^{kil}(\Lambda_U)_{lpn}(R_1)^{nm}_j(R_4)^p_{km}
\nonumber\\&&-6(\Lambda_D)^{lik}(Y_d)_{nl}(R_1)^{nm}_j(R_9)_{km}-3(\Lambda_E)_{kmj}(\Lambda_D)^{pmq}(R_5)^{ik}(R_7)_{qp}
\nonumber\\&&-3(Y_e)_{jk}(Y_d)^{qp}(R_5)^{ik}(R_7)_{qp}+6(\Lambda_D)^{lik}(Y_u)_{kp}(R_3)^{pn}_l(R_5)_{jn}
\nonumber\\&&+12C_L(R_1)^{kl}_j(R_1)^i_{kl}-2C_{H_2}(R_5)^{ik}(R_5)_{jk}
\nonumber\\&&-2C_{e^c}(R_5)^{ik}(R_5)_{jk}-3(R_1)^{lk}_j(R_1)^i_{mk}(\gamma_Q)^m_l
\nonumber\\&&-3(R_1)^{kl}_j(R_1)^i_{km}(\gamma_U)^m_l-(R_5)^{ik}(R_5)_{jm}(\gamma_E)^m_k
\nonumber\\&&-(R_5)^{ik}(R_5)_{jk}(\gamma_{H_2})
\end{eqnarray}

\begin{eqnarray}
(16\pi^2)^2(\beta^{(2)}_{m^2_{e^c}})^i_j&=&(16\pi^2)^2\beta{}_{m^2_{e^c}}^{(2)MSSM}
-6(Y_e)_{jk}(Y_e)^{qp}(R_5)^{ik}(R_7)_{pq}
\nonumber\\&&-6(\Lambda_E)_{jmk}(\Lambda_D)^{qmp}(R_5)^{ki}(R_7)_{pq}-6(Y_e)_{mj}(\Lambda_D)^{qmp}(R_2)^{i}(R_7)_{pq}
\nonumber\\&&-6C_{d_c}(R_3)^{ki}_l(R_3)^l_{kj}-4C_{H_2}(R_2)^{i}(R_2)_{j}
\nonumber\\&&-4C_{H_2}(R_5)^{ki}(R_5)_{kj}-6C_{u^c}(R_3)^{ki}_l(R_3)^l_{kj}
\nonumber\\&&-4C_L(R_7)^{li}(R_7)_{kj}-4C_{H_1}(R_2)^{i}(R_2)_{j}
\nonumber\\&&-2(R_5)^{ki}(R_5)_{mj}(\gamma_L)^m_k-2(R_2)^{i}(R_2)_{j}(\gamma_{H_1})
\nonumber\\&&-3(R_3)^{ki}_l(R_3)^l_{mj}(\gamma_U)^m_k-2(R_2)^{i}(R_2)_{j}(\gamma_{H_2})
\nonumber\\&&-3(R_3)^{ki}_l(R_3)^m_{kj}(\gamma_D)^l_m-2(R_5)^{ki}(R_5)_{kj}(\gamma_{H_2})
\end{eqnarray}

\begin{eqnarray}
(16\pi^2)^2\beta^{(2)}_{m^2_1}&=&(16\pi^2)^2\beta{}_{m^2_{1}}^{(2)MSSM}
-6(Y_d)^{lk}(Y_u)_{ln}(R_5)^{mn}(R_5)_{mk}-6(Y_e)^{lk}(\Lambda_D)_{pln}(R_9)^{nm}(R_3)^p_{mk}
\nonumber\\&&-6(Y_d)^{kl}(\Lambda_D)_{lpn}(R_9)^{nm}(R_1)^p_{km}+6(Y_d)^{kl}(Y_d)_{nl}(R_9)^{nm}(R_9)_{km}
\nonumber\\&&-3(Y_e)_{mk}(\Lambda_D)^{qmp}(R_2)^{k}(R_7)_{pq}+(Y_e)_{mk}(\Lambda_D)^{qpm}(R_2)^{k}(R_5)_{pq}
\nonumber\\&&-(Y_e)_{mk}(Y_e)^{mq}(R_2)^k(R_2)_q-6(Y_d)^{kl}(Y_u)_{kp}(R_3)^{pn}_l(R_2)_n
\nonumber\\&&+12C_{H_1}(R_9)^{kl}(R_9)_{kl}-2C_{H_2}(R_2)^{k}(R_2)_{k}
\nonumber\\&&-2C_{e^c}(R_2)^{k}(R_2)_{k}-3(R_9)^{lk}(R_9)_{mk}(\gamma_Q)^m_l
\nonumber\\&&-(R_9)^{kl}(R_9)_{km}(\gamma_U)^m_l-(R_2)^{k}(R_2)_{k}(\gamma_{H_2})
\nonumber\\&&-(R_2)^{k}(R_2)_{k}(\gamma_E)^m_k
\end{eqnarray}

\begin{eqnarray}
(16\pi^2)^2\beta^{(2)}_{m^2_2}&=&(16\pi^2)^2\beta{}_{m^2_{2}}^{(2)MSSM}
+12(Y_u)^{kl}(\Lambda_U)_{lpn}(R_7)^{mn}(R_4)^p_{km}+6(Y_u)^{kl}(Y_u)_{nl}(R_7)^{mn}(R_7)_{km}
\nonumber\\&&+12C_{H_2}(R_7)^{kl}(R_7)_{kl}+4C_{H_2}(R_2)^{k}(R_2)_{k}
\nonumber\\&&+4C_{H_2}(R_5)^{kl}(R_5)_{kl}-(R_2)^{l}(R_2)_{m}(\gamma_E)^m_l
\nonumber\\&&-3(R_7)^{kl}(R_7)_{km}(\gamma_D)^m_l-(R_2)^{k}(R_2)_{k}(\gamma_{H_1})
\nonumber\\&&-3(R_7)^{lk}(R_7)_{mk}(\gamma_Q)^m_l-(R_5)^{kl}(R_5)_{km}(\gamma_E)^m_l
\nonumber\\&&-(R_5)^{lk}(R_5)_{mk}(\gamma_L)^m_l
\end{eqnarray}

\begin{eqnarray}
(16\pi^2)^2(\beta^{(2)}_{m^2_R})^i&=&(16\pi^2)^2(\beta^{(2)RPV}_{m^2_R})+
6(\Lambda_D)^{lik}(\Lambda_D)_{lpn}(R_9)^{nm}(R_1)^p_{km}-6(\Lambda_D)^{lik}(Y_d)_{nl}(R_9)^{nm}(R_5)_{km}
\nonumber\\&&-3(Y_e)_{mk}(\Lambda_D)^{qmp}(R_5)^{ik}(R_7)_{pq}+6(\Lambda_D)^{lik}(Y_u)_{kp}(R_3)^{pn}_l(R_2)_n
\nonumber\\&&+12C_L(R_9)^{kl}(R_1)^i_{kl}-2C_{H_2}(R_2)^{k}(R_2)_{k}
\nonumber\\&&-2C_{e^c}(R_5)^{ik}(R_2)_{k}-3(R_9)^{kl}(R_1)^i_{km}(\gamma_U)^m_l
\nonumber\\&&-3(R_9)^{lk}(R_1)^i_{mk}(\gamma_Q)^m_l-(R_5)^{ik}(R_2)_{k}(\gamma_{H_2})
\nonumber\\&&-(R_5)^{ik}(R_2)_{m}(\gamma_E)^m_k
\end{eqnarray}

where
\beqn
&C_{Q}= \frak{4}{3}g_3^2 + \frak{3}{4}g_2^2 +\frak{1}{60}g_1^2,\quad
C_{u^c}= \frak{4}{3}g_3^2 +\frak{4}{15}g_1^2,\quad
C_{d^c}= \frak{4}{3}g_3^2 +\frak{1}{15}g_1^2,\cr
&C_{e^c}= \frak{3}{5}g_1^2,\quad
C_H = \frak{3}{4}g_2^2 +\frak{3}{20}g_1^2.\cr
\eeqn

We have the following results for the $\phi\phi$-type terms:

\beqn
(16\pi^2)^2\beta^{(2)}_{m_3^2}&=&(16\pi^2)^2\beta_{m_3^2}^{(2)MSSM}+
\eeqn
\beqn
(16\pi^2)^2(\beta^{(2)}_{m_K^2})^i &=&(16\pi^2)^2(\beta^{(2)RPV}_{m_K^2})+
\eeqn

In our calculation we use the results of  Ref~\cite{a21} which gives two-loop $\beta$-functions for a general gauge susy theory for standard and non-standard  soft susy braking terms , and also  we assume $m_F$  is zero in Eq.~(4)\cite{a21}(therefor $m_4$ and $m_r$ are zero in Eqs. ~(14) and (15)).

To check  our results,first we have calculated  one loop RGEs for both standard and non-standard terms, and have compared with ref~\cite{a12}. Our results are the same as theirs. Moreover;
to test the method  we  have obtained the two loop standard soft braking $\beta$-functions
which are consistent with results in ref~\cite{a22}.

\section{Conclusion}

In this paper we have expanded the study of the RG evolution of non-standard soft terms up two loop, and
presented the two-loop renormalisation of the R-parity violating extension of the MSSM
with the  most general possible set of soft breaking terms consistent with naturalness.

Typically, we expect effects of the two loop $\beta$-functions make a difference of several percent in compare with effects of one loop on the standard running analysis such as Higgs physics and the scalar quark sector of the MSSM; however it is quite difficult to make consequential estimates of the size of the two-loop corrections without committing to a specific model. Moreover it is desirable from the point of view of consistency to use the full set of $\beta$-functions

\end{document}